\documentclass[11pt]{article}
\usepackage{jheppub}
\usepackage{amsfonts}
\usepackage{dsfont}

\usepackage{amsmath}
\usepackage{url}
\usepackage{hyperref}
\usepackage{slashed}
\usepackage{tikz}
\usetikzlibrary{decorations.pathmorphing}
\usetikzlibrary{matrix}
\usepackage{graphicx}
\usepackage{epstopdf}
\usepackage{subfigure}
\usepackage{pgfplots}
\usepackage{amssymb}
\usepackage{color}
\usepackage{mathrsfs}
\usepackage{fancybox}
\usepackage{lipsum}
\usepackage{eurosym}
\usepackage{tcolorbox}
\def\C{{\mathbb C}}
\def\R{{\mathbb R}}
\usepackage{tikz,pgfplots}
\usetikzlibrary{decorations.pathmorphing}
\tikzset{snake it/.style={decorate, decoration=snake}}
\usepackage{simpler-wick}
\def\spc{\hspace{1pt}}

\usetikzlibrary{trees}
\usetikzlibrary{shapes.misc}
\usepackage{tkz-euclide}
\usepackage{verbatim}

\newcommand{\Del}{\Delta}

\newcommand{\be}{\begin{equation}}
\newcommand{\ee}{\end{equation}}

\def\[{\left [}
\def\]{\right ]}

\def\phys{}
\def\J{{\cal J}}

\numberwithin{equation}{section}

\usepackage{environ}
\NewEnviron{myequation}{%
\begin{eqnarray}
\scalebox{1.05}{$\BODY$}
\end{eqnarray}
}
\usepackage{dsfont}

\newcommand{\beq}{\begin{equation}}
\newcommand{\eeq}{\end{equation}}
 
\newcommand{\bea}{\begin{eqnarray}}
\newcommand{\ea}{\end{eqnarray}}
\newcommand{\barr}{\!\begin{array}}
\newcommand{\earr}{\end{array}\!}

\def\dd{{\mathrm d}}

\def\th{\theta}

\def \Oc {\mathcal{O}}
\def \bra {\langle}
\def \ket {\rangle}

\usepackage[natbib=false,maxbibnames=99,
    sorting=none,
    backend=biber,
    style=numeric,
]{biblatex}
\def\spc{\hspace{1pt}}

\counterwithout{equation}{section}
\def\nspc{{\hspace{-.75pt}}}

\def\xx{{\rm x\smpc}}

\renewcommand{\Large}{\large}

\def\be{\begin{equation}}
\def\ee{\end{equation}}

\def\la{\langle}
\def\bea{\begin{eqnarray}}
\def\eea{\end{eqnarray}}
\def\is{\!  & \!  = \!  &  \!}
\def\ra{\rangle}
\def\half{{\textstyle{\frac 12}}}

\def\ba{\bea}
\def\ea{\eea}

\def\be{\bea}
\def\ee{\eea}

\def\ra{\rangle}
\def\la{\langle}

\renewcommand\large{\fontsize{13}{13.5}\selectfont}

\def\spc{\hspace{1pt}}
\def\is{ &   =  & }

\def\nspc{\!\spc}

\def\l{\left}
\def\r{\right}
\def\darkgreen{green!50!black}
\def\dS{{\rm dS}}
\def\xx{\tau}

\begin{document}

\begin{flushright}
OU-HET-1309\\RIKEN-iTHEMS-Report-26
\end{flushright}
\vspace*{0.5cm}
\addtolength{\abovedisplayskip}{.8mm}
\addtolength{\belowdisplayskip}{.8mm}

\addtolength{\parskip}{.8mm}
\vspace{-10mm}

\title{Generalized Free Fields in de Sitter from 1D CFT}
\author[1,2,3]{Kanato Goto,}
\affiliation[1]{Department of Physics, Princeton University, Princeton, NJ 08544, USA}
\affiliation[2]{Department of Physics, The University of Osaka,
Machikaneyama-Cho 1-1, Toyonaka, Japan 560-0043}
\affiliation[3]{RIKEN Center for Interdisciplinary Theoretical and Mathematical Sciences (iTHEMS), RIKEN, Wako 351-0198, Japan}
\author[4,5]{Alexey Milekhin,} 
\affiliation[4]{Institute\! for\! Quantum Information\! and~Matter, California\! Institute\! of\! Technology, Pasadena, CA\! 91125,~USA}  
  \affiliation[5]{Department of Physics and Astronomy, University of Kentucky, Lexington, KY 40506, USA} 
\author[1]{Herman Verlinde,} 
\author[6]{Jiuci Xu}
\affiliation[6]{Department of Physics, University of California, Santa Barbara, CA 93106, USA}

\abstract{
We show that a pair of identical large $N$ 1D CFTs, like the low-energy limit of the SYK model or a line-defect inside a higher dimensional CFT, contains a natural sub-algebra of operators that comprise a generalized free field algebra living on a time-like geodesic in d+1-dimensional de Sitter spacetime. The construction uses large $N$ factorization, 1D conformal symmetry, and the split representation of de Sitter Green functions. We show that for 3D de Sitter spacetime, the holographic map extends into the bulk and reduces to the standard HKLL prescription adjusted to de Sitter spacetime. We  describe how our construction is automatically implemented in a covariant version of Schwarzian quantum mechanics and comment on the relevance of our results to the de Sitter/DSSYK correspondence.}

\date{\today}

\maketitle
\section{Introduction}
\addtolength{\baselineskip}{.25mm}

\def\half{\frac{\raisebox{-1pt}{\small 1}}{\raisebox{1pt}{\small 2}}}

There exist various proposals for a candidate holographic description of quantum de Sitter spacetime \cite{Strominger:2001pn,Banks:2006rx,Anninos_2012, Coleman:2021nor, susskind2022sitterspacedoublescaledsyk,narovlansky2023doublescaledsyksitterholography,Milekhin:2023bjv}. While each approach has booked some successes, it is fair to say that the subject is still in an early stage of development. A promising avenue that received some recent attention is the idea of worldline holography \cite{Anninos_2012,Chandrasekaran:2022cip, Witten:2023xze, narovlansky2023doublescaledsyksitterholography}, which aims to describe the observations of an idealized observer moving along a timelike trajectory inside de Sitter spacetime. To set up a theory of worldline holography, it is necessary to be able to take a limit where gravity is turned off or weak enough to allow for a controlled semi-classical treatment. Ideally, copying the successful large $N$ approach to AdS/CFT \cite{Leutheusser:2022bgi}, there should exist an operator sub-algebra in the dual quantum theory that generates a generalized free field algebra \cite{banks1998adsdynamicsconformalfield, Hamilton:2006az} of local QFT observables accessible to the de Sitter observer. 

It is then a natural question to ask: Does there exist a natural class of quantum systems that contain a generalized free field algebra on the geodesic worldline of a de Sitter observer? In this note we point out that such a class of quantum systems indeed exists in the form of a coupled system of two one-dimensional large $N$ CFTs (labeled by L and R) subject to the zero energy constraint \cite{Anninos_2012,narovlansky2023doublescaledsyksitterholography}
\bea
\label{physconstraint}
(H_L - H_R) |\Psi_{\phys}\rangle \is 0 
\eea
One can view the above constraint as the statement that one of the CFT$_1$'s acts like a clock for the other CFT$_1$, c.f. \cite{Chandrasekaran:2022cip}. For a given microscopic realization at large but finite $N$ one can consider two versions of this condition: either one can view \eqref{physconstraint} as  implementing a one-to-one pairing between energy eigenstates of the two systems or a coarse grained condition that only becomes exact upon taking the large $N$ limit. For most of this note, we will adopt the latter definition. We will comment on the difference between two implementations of the energy constraint in section \ref{sec:fine-coarse}
and Appendix~\ref{sec:appendix}.

If we impose the condition \eqref{physconstraint} as a definition of physical states, then physical operators $\mathbb{O}_\Delta(\xx)$ must be relational observables that commute with this constraint \cite{narovlansky2023doublescaledsyksitterholography} 
\bea
\label{eq:physop}
[H_L\!-\nspc H_R,{\mathbb{O}_{\Delta}}] \is 0.
\eea
A natural class of such observables are given by a convolution product of local scaling operators ${\cal O}^L_{d/2-\Delta}$ and ${\cal O}^R_{\Delta}$.
Following this prescription, we will construct a class of generalized free operators that satisfy the constraint \eqref{physconstraint} and show that their Wightman correlation functions coincide 
with those of a free massive scalar field $\phi(\xx)$ placed along a given time-like geodesic in $d+1$-dimensional de Sitter spacetime
\bea
\label{genfreefield}
\boxed{ \ \Bigl\langle {\mathbb{O}}_\Delta(2\xx_{1}) \ldots {\mathbb{O}}_\Delta(2\xx_{n})\Bigr\rangle_{\rm{CFT}_1} = \; \Bigl\la \phi(\xx_1) \ldots \phi(\xx_n) \Bigr\ra_{\!\rm dS_{d+1}} {}^{\strut}_{\strut}}
\eea
as a function of the geodesic time $\tau_i$ along the geodesic.
Here the mass of the scalar field is related to the CFT$_1$ scale dimension $\Delta$ via the formula 
\bea
\label{eq:Delta}
m^2 = 4\Delta(d/2-\Delta).
\eea
The identity \eqref{genfreefield} is the main result of this note. It is not an entirely new insight and a similar statement has been noted before in \cite{Anninos_2012,narovlansky2023doublescaledsyksitterholography}. 
This universal result, however, does not appear to be widely appreciated and given the recent interest in de Sitter worldline holography, it seems worthwhile to present it here. 
The relation \eqref{genfreefield} for $d=2$ also plays an essential role in the proposed duality between DSSYK and 2+1 de Sitter gravity \cite{narovlansky2023doublescaledsyksitterholography}.
Within this $d=2$ context, our longer term aim is to go beyond the symmetry arguments of \cite{Anninos_2012} and provide a microscopic realization of de Sitter holography, paving the road to a possible full dictionary, including higher-point functions, backreaction and thermodynamics. 

Our construction relies on only three physical inputs: 1D conformal symmetry, large $N$ factorization, and the split representation of de Sitter Green functions, see e.g. \cite{Sleight:2019mgd,Xiao:2014uea}. As a concrete example of a 1D large $N$ CFT, we will consider the large $q$ limit of the SYK model at finite temperature. Other 1D CFTs, such as the low energy limit of 1D conformal defects inside a higher dimensional CFT, are also suitable examples. Our main CFT$_1$ input is that it possesses a set of scaling operators ${\cal O}_\Delta(\xx)$ with thermal Wightman two-point functions
\beq
\label{twopt}
G_\Delta(t) = \bigl\langle {\cal O}_\Delta(t) {\cal O }_\Delta(0) \bigr\rangle  =e^{-i \pi \Delta} \biggl(\frac 1{2 \sinh\bigl(t/2-i\epsilon\bigr)} \biggr)^{\!2\Delta}
\eeq
Here we chose units such that the inverse temperature $\beta = 2\pi$.

In the following sections, we first define the physical operators $\mathbb{O}_\Delta$ and show that its two point function matches the de Sitter Green function. Next we give a geometric explanation of how our construction works in terms of the so-called split representation of the de Sitter Green function, or perhaps more physically, via Huygens principle or Green's theorem. We then explain how to implement large $N$ factorization and reproduce Wick's theorem. In section~\ref{sec:matrix}, we specialize to 2+1D de Sitter spacetime and give a group theoretic treatment and make contact with the HKLL prescription for constructing bulk operators in dS/CFT. We conclude with some remarks on the difference between the fine-grained and coarse-grained equal energy constraint and on the relevance of these results to the de Sitter/SYK correspondence. In the Appendix, we point out that the physical operators introduced in this note naturally appear in a reparametrization invariant version of Schwarzian quantum mechanics \cite{Maldacena:2016hyu,Jensen:2016pah,maldacena2025dimensionalnearlysittergravity,Mertens:2017mtv} and fill in more details of the group theoretic perspective on the physical operators.

\section{De Sitter Green function as a CFT$_1$ two-point~function}

Let ${\cal O}^L_{\Delta}(t)$ and ${\cal O}^R_{\Delta}(t)$ denote two  scaling operators with thermal two point functions given in equation \eqref{twopt} in two decoupled 1D CFTs.  We now introduce the following class of physical operators that commute with the equal energy constraint \eqref{physconstraint}
\bea
\label{physoperator}
\boxed{\ \ \mathbb{O}_\Delta(\tau) = {\cal N} \int\! dt\; {\cal O}^L_{d/2-\Delta}(t) {\cal O}^R_{\Delta}(\tau-t)_{\strut}^{\strut}\ \ }
\eea
Here the prefactor ${\cal N}$ in \eqref{physoperator} is an infinite renormalization factor defined via
\beq
\label{tlimit}
{\cal N} \int\! dt\, \equiv\,  \lim_{T\to \infty}\frac{1}{\sqrt{2T}} \int_{-T}^{T}\!\!\!dt
\eeq
The infinite renormalization is necessary so that the correlation functions of the physical operators remain finite. The integral over $t$ can be understood as a noncompact $O(1,1)$ group
average, enforcing invariance under $H_L - H_R$, with $\mathcal N$
removing the infinite group volume.\footnote{There is a potential ambiguity in the definition \eqref{tlimit}, namely whether the  $T\to \infty$ limit is taken before or after taking the large $N$ limit, or equivalently, whether the integral in \eqref{physoperator} imposes a microscopic or coarse grained equal energy constraint. We will comment this freedom and how it shows up in our results in the concluding section \ref{sec:fine-coarse} and in Appendix A. In the main text we adopt the coarse grained definition that the equal energy constraint is imposed after taking the large $N$ limit. An additional subtlety is that the limit \eqref{tlimit} has to be taken simultaneously, with the same value of $T$, for all physical operators in a given correlator.} We will later apply an additional finite renormalization, so that the correlation functions match with those of the generalized free field in dS$_{d+1}$. 
The class of operators  \cite{Anninos_2012,narovlansky2023doublescaledsyksitterholography}  manifestly commute with the zero energy constraint \eqref{physconstraint}.
Here $d$ is an a priori free parameter, but we will choose it to be an integer; it will determine the number of space dimensions of the dual de Sitter spacetime.

Our goal is to compute the two-point function 
\bea
\label{gphys}
\mathbb{G}_\Delta(\tau) = \bigl\langle \mathbb{O}_\Delta(\tau) \mathbb{O}_\Delta(0) \bigr\rangle
\eea
and compare with the known expression for the de Sitter Green function. The two point function \eqref{gphys} is given by the convolution product
\beq
{\mathbb{G}_{\Delta}(\tau) = \int {dt}\ G_\Delta(\tau-t) G_{d/2-\Delta}(t)}.
\eeq
of two CFT$_1$ two-point functions \eqref{twopt}.
Below we will evaluate this convolution product both in the position and frequency domain. We start with the latter.

Fourier transforming the CFT$_1$ two-point function \eqref{twopt}~yields\footnote{
Here and in what follows, we often use the notation $f(a\pm b) = f(a+b)f(a-b)$.}
\beq
\tilde{G}_\Delta(\omega) =   e^{{{\pi}\omega}} \, \frac{\Gamma({ \Delta \pm i\omega})}{\Gamma(2\Delta)}
\eeq
The two-point function $\mathbb{G}_\Delta(\tau)$ of two physical operators \eqref{physoperator} is found by taking the inverse Fourier transform of
\bea
\tilde{\mathbb{G}}_{\Delta}(\omega) \is\, {\tilde{G}_{\Delta}(\omega) \tilde{G}_{d/2 -\Delta}(\omega)}\notag\\[-1.5mm]\\[-1.5mm]\notag
\is    e^{2\pi\omega} \,  \frac{\Gamma ( \Delta \pm i\omega)
\Gamma (d/2-\Delta \pm i\omega)}{\Gamma(2\Delta) \Gamma(d-2\Delta)}
\ \ 
\eea
This inverse transform can be explicitly computed by closing the integration contour in the lower-half plane, picking up the poles of $\Gamma \l( \Delta - {i \omega}\r) \Gamma \l( d/2 -\Delta  - {i \omega}\r)$.  
This yields a sum of hypergeometric functions,
which after some rearrangement can be shown to equal the Wightman Green function 
of a scalar field with $m^2=4\Delta(d/2-\Delta)$ in d+1-dimensional de Sitter spacetime 
as a function of the proper time difference $\tau$ 
\bea
\boxed{\ 
\mathbb{G}_{\Delta}(\tau ) = \mathcal{N}_{d,\Delta}  {G}^{\dS}_{\Delta}(\tau/2) \ \scriptsize {}^{\strut}_{\strut} }_{\strut}
\eea
\vspace{-7mm}

\noindent
with\\[-7mm]
\bea
\mathcal{N}_{d,\Delta} = \frac{\pi^{3/2+d/2}}{2^{d-2}} \frac{(d+1) \Gamma(d) }{ \Gamma(3/2+d/2)\Gamma(d-2\Delta)\Gamma(2\Delta) } 
\eea
From now on, we will absorb a factor of $1/\sqrt{\mathcal{ N}_{d,\Delta}}$ in our definition of the physical operators.

The explicit de Sitter Green's function reads
\begin{eqnarray}
\label{eq:gexplicit}
G^{\rm dS}_\Delta(\tau)\is c_{\Delta,d} \; { }_2 F_1\nspc\biggl(2\Delta,d-2\Delta ;\frac{d+1}2; \sigma\biggr)
\end{eqnarray}
where $\Delta,d/2-\Delta$ are the two solutions to \eqref{eq:Delta} and
\beq
\label{distanceone}
\sigma = \frac {1+\cosh(\tau)}2, \qquad c_{\Delta,d} = \frac{\Gamma(2\Delta)\Gamma(d-2\Delta)} {(4 \pi)^{\frac{d+1} 2}\Gamma\bigl(\frac{d+1}{2}\bigr)}
\eeq
It is remarkable result that this Green function can be mapped in such a simple way to the two-point function of the physical operators \eqref{physoperator} in the doubled CFT$_1$. As explained in \cite{Anninos_2012} and below, the correspondence can be naturally explained via the fact that d+1-dimensional de Sitter spacetime has a hidden SU$(1,1)$ symmetry.

\medskip

\section{Split representation of the de Sitter Green function}

\vspace{-1mm}

We now present a geometric explanation of the above construction in terms of the so-called split representation of the de Sitter Green function, see e.g. \cite{Sleight:2019mgd}. The physical intuition is that the bulk-to-bulk Green function in de Sitter spacetime can be decomposed as a convolution of two bulk-to-boundary Green functions. This decomposition may be viewed as a version of Green's theorem or Huygens' principle, as illustrated in figure 1. 

Let us introduce flat coordinates $(\eta,\vec{x})$ in which the de Sitter metric takes the form 
\begin{equation}
\dd s^2 =\frac{1}{\eta^2}\bigl(-\dd\eta^2+\dd\vec x^{\,2}\bigr).
\end{equation}
In this coordinate system, the geodesic distance between two bulk points $X= (\eta,\vec{x})$ and $X' = (\eta',\vec{x}')$ in de Sitter spacetime is expressed as
\bea
\label{distancetwo}
\sigma  = 1 - \frac{(\vec{x}\!-\!\vec{x}')^2\!\!-\!(\eta\!-\!\eta')^2\!\!\!}{{4 \eta \eta'}}. 
\eea
The flat coordinate system is particularly well suited for the analysis of boundary behavior 
near future infinity $\mathscr I^+$. As $\eta'\to0^{+}$, the Green's function  behaves as
\bea
G^{\rm dS}_\Delta(X,X') \; \to\;  \eta'^{\,2\Delta}K_\Delta(X;\vec x') + \eta'^{\,d-2\Delta}K_{d/2-\Delta}(X;\vec x')
\label{eq:asymp}
\eea
The two independent falloffs correspond to the conjugate pair $\Delta_\pm$ that play a key role in the construction of physical operators \eqref{physoperator}.
In flat slicing, the bulk--to--boundary propagator takes the explicit form
\bea
K_\Delta(X;\vec x') \is c_\Delta e^{-2\pi i \Delta}\biggl(\frac{\eta}{\eta^2 -(\vec x-\vec x')^2}\biggr)^{2\Delta}\!, \ \   c_\Delta =  \frac{\pi^{-1-d/2}}{4^{2 \Delta+1}} \Gamma(d/2\nspc-2\nspc \Delta)\Gamma(2 \Delta)
\label{eq:K} 
\eea
For principal series representations one has $c_{d/2-\Delta}=c_\Delta^*$.
The appearance of both kernels $K_\Delta$ and $K_{d/2-\Delta}$ anticipates the reflection symmetry that will emerge below.
 
The near--boundary structure \eqref{eq:asymp} implies that bulk propagation can be expressed entirely in terms of boundary data. This leads to a split representation of the bulk Green function, see e.g. \cite{Sleight:2019mgd}. Applying Green's theorem to two solutions of the Klein--Gordon equation while substituting the asymptotic expansion \eqref{eq:asymp}, all terms involving identical falloffs cancel. After a bit of algebra, one finds that the remaining terms combine~into~\cite{Sleight:2019mgd} 
\bea
G^{\rm dS}_\Delta(X,Y) =  4\nu \sin(\pi\nu) \int_{\mathscr{I}^+ \text{or} \mathscr{I}^-}\!\!\!\!\!\!\!\dd^d x'\, K_\Delta(X;\vec x')K_{d/2-\Delta}(\vec x',Y),
\label{eq:split}
\eea
with $\nu =d/2- 2 \Delta$.
 This split representation makes explicit that bulk propagation involves a pairing of conjugate boundary modes.

  \begin{figure}[t]
  \centering
\begin{minipage}{0.5\textwidth}
\centering    \includegraphics[scale=.82]{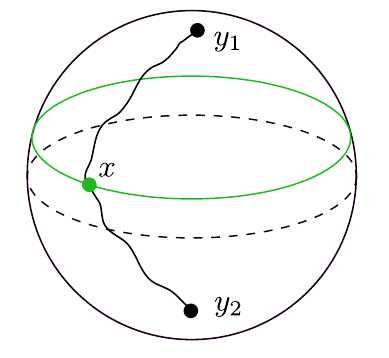}
  \end{minipage}
\begin{minipage}{0.48\textwidth}
  \centering    \includegraphics[scale=1.4]{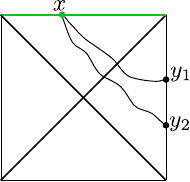}
\end{minipage}
\caption{Euclidean (left) and Lorentzian (right) setup of the Green's theorem for de Sitter.    In Lorentzian de Sitter, the Huygens surface (\textcolor{\darkgreen}{green}) is placed at (past or) future infinity. 
}
    \label{fig:sphere}
\end{figure}

We wish to relate the split representation \eqref{eq:split} of the Green function to the definition \eqref{physoperator} of our physical observables . Without loss of generality, let us consider the Green function between two special bulk-points $X\!= \!(\eta,\vec{x})$ and $Y\!=\!(\eta',\vec{y})$ with  $\vec{x}=\vec{y}$.  The integral in \eqref{eq:split} then only depends on the radial variable $r=|\vec{x}-\vec{x}'|$. Hence we can trivially integrate over the angular variables, reducing the integral over $\mathscr{I}^\pm$ to 
\footnote{{The proper $i \epsilon$ prescription takes $\eta \rightarrow \eta^{-i \epsilon}$ and $r^2-1$ to $r^2-e^{-2 i \epsilon}$.}}
\ba
\label{gintegral}
G^{\rm dS}_{\Delta}(\tau )\is C_\Delta \int_0^\infty\!\!\!\! dr\, r^{d-1}\Bigr(\frac{\eta}{\eta^2\!-r^2}\Bigr)^{2\Delta}(r^2\nspc -\nspc 1)^{2\Delta-d}\, ,\notag\\[-1.5mm]\\[-1.5mm]\notag
&& C_\Delta = \frac{(4\Delta\nspc -\nspc d)\Gamma(2\Delta) \Gamma(d\nspc -\nspc 2\Delta)}{(4\pi)^{d/2 +1}}
\ea 
Here for convenience we put $\eta'=1$ and $\tau$ denotes the geodesic distance between $X$ and $Y$. Using \eqref{distanceone} and \eqref{distancetwo} we find that $\eta = e^\tau$. Equation \eqref{gintegral} can be recognized as the integral representation of the hypergeometric function \eqref{eq:gexplicit}.

Let us compare the integral expression \eqref{gintegral}  with the two-point function  $\mathbb{G}_{\Delta}(\tau )$
of the physical operators \eqref{physoperator}. It will be helpful apply a conformal mapping $z= e^t$ and write  
\bea
\mathbb{O}^{\phys}_{\Delta}(\tau)\is
e^{2\tau \Delta}{\cal N}\int_0^\infty\!\!\!\!\frac{ dz\,}{z^{1-d/2}} \, {\mathcal{O}}^L_{d/2-\Delta}\left(z\right)\mathcal{O}^R_{\Delta}\left(e^{2\tau} z\right)\ \ 
\eea
The  two-point function $\mathbb{G}_{\Delta}(\tau )$
 is then given by a double integral  
\bea\label{Gphys}
e^{2\tau \Delta} {\cal N}^2\int_0^\infty\!\!\!\int_0^\infty\!\frac{ dzdz'} {(zz')^{1-d/2}\!\!\!\!}\;\;\, \langle  \mathcal{O}_{\Delta}^R(z)\mathcal{O}_{\Delta}^R(e^{2\tau} z')\rangle \langle \mathcal{O}_{d/2-\Delta}^L(z)\mathcal{O}_{d/2-\Delta}^L(z')\rangle
\eea
Plugging in the standard expression for the conformal two point function in flat coordinates
\ba
\langle \mathcal{O}_{\Delta}(z)\mathcal{O}_{\Delta}(z')\rangle=\frac{1}{(z'-z)^{2\Delta}}\, .
\ea
and changing the integration variable $z'$ to $\bar{z}=1/z'$, we find that the integrand only depends on a single radial variable $r=(z\bar{z})^{1/2}$. 
Writing 
$z=re^{u},\bar{z}=re^{-u}$ and performing the {\it trivial} integration over $u$ 
yields an integral expression  that looks identical to \eqref{gintegral}  upto an overall constant factor.
Thus we  again confirm the two-point function of the physical observables reproduces the Green function in the de Sitter spacetime.
\medskip

\section{Wick's theorem}
\vspace{-1mm}

To get a more complete holographic dictionary, we need to identify a notion of large $N$ factorization such that $2n$-point functions decompose into a Wick sum of products of two-point functions
\bea
\ \ \bigl\la \mathbb{O}_1 \, .\; .\ .\ \mathbb{O}_n \bigr\ra = \la \mathbb{O}_1 \mathbb{O}_2\ra \ldots\la \mathbb{O}_{n-1} \mathbb{O}_n\ra +  {\rm permutations}
\label{wickrule}
\eea
Operators that satisfy this rule are known as generalized free fields \cite{banks1998adsdynamicsconformalfield,Hamilton:2006az,Leutheusser:2022bgi}. Realizing this factorization formula in our setting is a bit more subtle than usual, since the $\mathbb{O}_i$'s are not local CFT operators but non-local convolution products. Here we outline a possible implementation in the context of the SYK model.

Consider two identical SYK models with Hamiltonians \cite{kitaev2015simple,Maldacena_2016}
\bea
H_{L} =\!\! \sum_{i_1,...,i_p}  J_{i_1,..i_p} \psi_{i_1}^{L} \psi_{i_2}^{L}\ldots \psi_{i_p}^{L},
\quad \quad
H_{R} = \sum_{i_1,...,i_p}  J_{i_1,..i_p} \psi_{i_1}^{R} \psi_{i_2}^{R}\ldots \psi_{i_p}^{R}
\eea
with identical gaussian random couplings with variance 
\bea
\overline{J_{i_1\dots i_p}J_{i_1\dots i_p}} 
= \frac{2^p N \J^2 }{2p^2 \Bigl(  {{N}\atop p}  \Bigr)}
\eea
The SYK model is exactly solvable in the double scaling limit at large $N$ and large~$p$ with $\lambda=p^2/N$ finite \cite{Berkooz:2018jqr}. Here we will only consider the strict large $N$ limit with $\lambda\to 0$. The  two point function of  scaling operators  ${\cal O}_\Delta(t)$ in this limit takes the standard CFT$_1$ form \eqref{twopt} with $t = \frac{2\pi v}{\beta} t_{12} + i \pi (v-1)$  (here ${\cos \frac {\pi v} 2}= \frac{\pi v}{\beta}$) a suitably rescaled and shifted relative time coordinate \cite{Maldacena_2016,Lin:2023trc} (see also section 6).  Hence we can apply the same derivation as given above to show that the two-point function of two physical operators \eqref{physoperator} match with the Green function along a geodesic in $d+1$-dimensional de Sitter spacetime. This result generalizes the correspondence found in \cite{narovlansky2023doublescaledsyksitterholography} for the case of $2+1$-D de Sitter to arbitrary dimensions.

To construct a set of generalized free field operators that satisfy Wick's theorem we now define the product of the $L$ and $R$ scaling operators as follows \cite{narovlansky2023doublescaledsyksitterholography}
\bea
\label{newphys}
\mathbb{O}_\Delta(\tau) = \int \! dt \, {\cal O}^L_{\Delta}(t) {\cal O}^R_{d/2-\Delta}(\tau-t) \, \equiv\, \int \! dt \;  k^{IJ} \, \psi^L_{I}(t) \psi^R_{J}(\tau-t)
\eea
Here $
 I = \{i_1,\ldots, i_r\}$ and $J = \{j_1,\ldots, j_s\}$ denote multi-indices of length $2 r= p\spc \Delta$ and $2s=p\spc(d/2 -\Delta)$ and $
\psi^L_{I}$ and $\psi^R_J$ are short-hand for $\psi_{i_1}^L\ldots \psi_{i_r}^L$ and $\psi_{j_1}^R\ldots \psi_{j_s}^R.$ The couplings $k^{IJ}$ are random numbers picked from a suitably normalized gaussian ensemble with diagonal variance
\bea
\label{variance}
\bigl\langle k^{I\nspc J} k^{I'\!\nspc J'} \bigr\rangle =  \delta^{I\nspc I'}\delta^{J\nspc J'} {\,\frac{2^{r+s}N \kappa^2 }{2p^2 \Bigl(\!  {{N}\atop r} \nspc \Bigr)\Bigl(\!  {{N}\atop s}\nspc  \Bigr)}}
\eea 
The new ingredient here is that the couplings $k^{IJ}$ do {\underline{not}} factorize into a product of two separate random couplings for the left and right model. Since the variance \eqref{variance} vanishes unless $I = I'$ \underline{and} $J=J'$, performing the gaussian averaging over the couplings $k^{IJ}$ generates the Wick rule \eqref{wickrule}. Thus the physical operators of the form \eqref{newphys} constitute a set of generalized free fields that live on a geodesic worldline in dS$_{d+1}$. This statement holds in the strict large~$N$, $\lambda\to 0$ limit and can be straightforwardly generalized to other large $N$ CFT$_1$'s. We briefly comment on the extension of this result to finite $\lambda$ in the concluding section.

\def\darkgreen{green!50!black} 
\def\R{{\mathbb{R}}}

 \medskip

 \section{Physical operators as matrix elements}\label{sec:matrix}

From here on we specialize to the case of dS$_3$.  The above relations then admit a natural interpretation in terms of
representation theory of the de Sitter isometry group $SL(2,\mathbb{C})$. 3D de Sitter spacetime is also of special interest as a candidate holographic dual for the double scaled SYK model \cite{narovlansky2023doublescaledsyksitterholography,susskind2022sitterspacedoublescaledsyk}.

Thus far we restricted the coordinate $z = e^t$ to be positive. For the following discussion, we will extend the domain of the 1D CFT operators ${\cal O}_\Delta(z)$ to the full real $z$-axis.\footnote{We define ${\cal O}_\Delta(z)$ on the negative real axis $z<0$ via analytic continuation. This can be done in two ways: depending on how we choose to go around the branch cut, we either pick up a factor of  $e^{2\pi i \Delta}$ or $e^{-2\pi i \Delta}$. For simplicity of presentation, we will suppress this subtlety here in the main text. For more discussion of the $i\epsilon$-prescription, see Appendix B and C.}
The 1D conformal group then acts on the primary operators via Möbius transformations 
\bea
\label{eq:Mobius}
g\! \cdot \mathcal{O}_\Delta(z) & \equiv &
(c z+d)^{-2\Delta}\,
\mathcal{O}_\Delta\!\Bigl(\frac{a z+b}{c z+ d}\Bigr)\notag \\[-2mm]\\[-2mm]\notag
g\is \biggl(\mbox{$\begin{array}{cc} a& b \\[-.25mm]c & d  \end{array}$}\biggr) \in  {SL}(2,\mathbb{R})
\eea
It will be helpful to complexify this M\"obius group to $SL(2,\mathbb{C})$. 3D de Sitter spacetime is a coset manifold $SL(2,\C)/SU(1,1)$. We can thus associate to a bulk point $X = (\eta,x,\bar{x})$ in dS$_3$ a coset representative $g_X \in {SL}(2,\C)$ via 
\begin{equation}
a=\frac{\bar x}{\eta},
\quad
b= \eta - \frac{x\bar{x}}{\eta},  \quad  c=-\frac{1}{\eta},
\quad
d=\frac{x}{\eta}.
\label{eq:alphabeta}
\end{equation} 
The Möbius transformation \eqref{eq:Mobius} associated to this bulk point $X$ becomes
\bea
\label{groupact}
g_X\! \cdot\nspc \mathcal{O}_\Delta(z) = \Bigl(\frac{\eta}{z-x}\Bigr)^{2\Delta}\,\mathcal{O}_{\Delta}\!\left(-\bar x - \frac {\eta^2}{z-x}\right)\,
\eea
The unit element $g=\mathds{1}$ correspond to the base point $(\eta,x,\bar{x}) =(1,0,0)$.
In order for $g_X$ to be a proper 1D conformal transformation, we need to restrict $g_X$ to lie in the real M\"obius group $SL(2,\mathbb{R})$. This amounts to placing the bulk point $X$ on real dS$_2$ slice $x,\bar{x} \in \mathbb{R}$ within dS$_3$. Below we will restrict to this real slice.

The group action \eqref{groupact} points to the following natural generalization of our physical operator \eqref{physoperator} for a general bulk location on the real dS$_2$ slice of dS$_3$
\bea
\mathbb{O}^{\phys}_\Delta(X) = \int \! \frac{\dd z}{2\pi i}\,  \widetilde{\mathcal{O}}^{\,L}_{1-\Delta}\nspc(z)\,  g_X\!\cdot \mathcal{O}^{R}_\Delta(z) 
\label{eq:OphysSU11}
\eea 
\noindent
Here we define the integral runs over the full real $z$ axis.  This formula extends our original definition \eqref{physoperator} of the worldline observables beyond the special case when the location $X$ lies on the 1D static geodesic $x = \bar{x} = 0$ and $\eta = e^{\tau}$. Note that by introducing this more general class of physical operators, we have relaxed our worldline holography perspective to enable a more complete bulk reconstruction. In particular, the generalized physical operators 
$\mathbb{O}_\Delta(X)$ in \eqref{eq:OphysSU11} do not all preserve the same Hamiltonian constraint \eqref{physconstraint}.\footnote{The operator $\mathbb{O}_\Delta(X)$ is invariant under a 1D subgroup of the $SL(2,\mathbb{R}) \times SL(2,\mathbb{R})$ conformal group of the coupled CFT$_1$ generated by $g_X^\dag\otimes g_X$ acting on ${\cal O}^L_{1-\Delta}$ and ${\cal O}^R_{\Delta}$.}
Nonetheless, the group theoretic viewpoint yields some useful insights. 

To exhibit the geometrical meaning of \eqref{eq:OphysSU11}, let us write the dual scaling operator $\widetilde{\mathcal O}_{1-\Delta}^L$ as the 1D shadow transform of a scaling operator 
of weight $\Delta$ \cite{Simmons_Duffin_2014,Sun:2021thf}
\bea
  \label{shadowtransform}
\widetilde{\mathcal{O}}_{1-\Delta}(z) \equiv      {\cal S}_1 \mathcal{O}_\Delta(z) =        \int d\bar z \,\frac{{\mathcal{O}}_{\Delta}(\bar z)}{(z-\bar z)^{2(1-\Delta)}\!\!\!}\,\; .
\eea 
This 1D shadow transform $\mathcal{S}_1$ acts as an intertwiner between two conjugate representations of the 1D conformal group $SL(2,\mathbb{R})$ labeled by ${\Delta}$ and ${1-\Delta}$. Plugging this into \eqref{eq:OphysSU11} gives, after a bit of algebra, the following suggestive expression for the new physical operator $\mathbb{O}_\Delta(X)$
\bea
\mathbb{O}^{\phys}_\Delta(X) = \int_{\mathbb{R}^2} \! \frac{\dd z\spc\dd\bar{z}}{2\pi i}\, \l( \frac{\eta}{-\eta^2+(\bar{z}-\bar{x})(z-x)} \r)^{2(1-\Delta)}  {\mathcal{O}}^{\,L}_{\Delta}\nspc(\bar{z})\mathcal{O}^{R}_\Delta(z) 
\label{eq:Ophystwo}
\eea 
This formula motivates us to introduce the bi-local physical operator ${\mathbb{O}}_\Delta(z,\bar{z})$ given by the product of two CFT$_1$ scaling operator with the same scaling dimension $\Delta$
\bea
\label{ophyslocal}
{\mathbb{O}}_\Delta(z,\bar{z}) \, = \, {\mathcal O}_\Delta^L(\bar{z})\, {\mathcal O}^R_\Delta(z)
\eea
acting on the tensor product Hilbert space ${\mathcal H}_L\otimes {\mathcal H}_R$ of the left- and right CFT$_1$. We propose that the holographic dictionary maps ${\mathbb{O}}_\Delta(z,\bar{z})$  to a local operator on future infinity $\mathscr{I}^+$ of 3D de Sitter spacetime. 

 Equation \eqref{eq:Ophystwo} looks like a two-time integral over two real time-coordinates $z$ and $\bar{z}$. By using analyticity, we can promote $z$ and $\bar{z}$ to two independent complex coordinates on $\mathbb{C}^2$.  Let us now assume that we can perform a double Wick rotation to the real slice on which $z$ and $\bar{z}$ are complex conjugate variables.\footnote{This in effect assumes that we can treat each CFT$_1$ operator ${\cal O}$ as a chiral lorentzian CFT$_2$ operator and that both combined can be rotated into a single non-chiral euclidean CFT$_2$ operator ${\mathbb{O}}$. We will justify this assumption a posteriori by our subsequent analysis and the results of sections 2 and 3 and Appendix B and C.} The bi-local operator ${\mathbb{O}}_\Delta(z,\bar{z})$ then transforms as a primary operator with equal left and right scaling dimension $\Delta_L= \Delta_R = \Delta$ under the 2D conformal group $SL(2,\mathbb{C})$ 
\bea
\pi_\Delta(g)\cdot \mathbb{O}_\Delta(z, \bar{z})
= (c z+d)^{-2 \Delta}\,\left(\bar{c}  \bar{z}+\bar{d} \right)^{-2 \Delta}
\mathbb{O}_\Delta\left(\frac{a z+b}{c z+d}, \frac{\bar{a} \bar{z}+\bar{b}}{\bar{c} \bar{z}+\bar{d}}\right) .
\eea
Here $\pi_\Delta(g)$ denotes the action of $g \in SL(2,\mathbb{C})$ in the representation with weight $2\Delta$.
Group theoretically, we can thus view ${\mathbb{O}}_\Delta(z,\bar{z})$ as an operator valued vector in the corresponding representation space $\mathcal H_{2\Delta}$. Let $|\mathbb{O}_\Delta\rangle$ denote the vector in $\mathcal H_{2\Delta}$ such that
\bea
{\mathbb{O}}_\Delta(z,\bar{z}) \, = \, \bigl \langle  z,\bar{z}\,\bigr|\spc \mathbb{O}_\Delta  \bigr\rangle
\eea
We can then write the new physical operator \eqref{eq:OphysSU11}-\eqref{eq:Ophystwo} as an operator valued matrix element of $g_X$ in the representation $\mathcal H_{2\Delta}$ as follows
\bea
\label{eq:matrixelement}
\boxed{\  \ \mathbb{O}^{\phys}_\Delta(X) = \langle 0| \pi_\Delta(g_X) |\spc  \mathbb{O}_\Delta \spc\rangle \ \footnotesize {}^{\strut}_{\strut}\, }
\eea
This matrix element is manifestly invariant under left action of the $SU(1,1)$ stabilizer subgroup of the state $|0\rangle$ corresponding to the special point $(\eta,x,\bar{x}) =(1,0,0)$. We will unpack and study this formula below and in Appendix \ref{app:group}.

The representation \eqref{eq:matrixelement} of the physical operator $\mathbb{O}_\Delta(X)$ as a matrix element has several useful applications. First, it directly explains why the physical operators satisfy the bulk Klein--Gordon equation: the $SL(2,
\mathbb{R})$ quadratic Casimir acts functions of $g$ as the Klein-Gordon wave operator $\square$ in dS$_3$, and matrix elements in the representation $\mathcal H_\Delta$ \cite{harish1969harmonic,vilenkin1978special,Sun:2021thf} are eigenfunctions with eigenvalue 
\be
\mathcal C_2 = 4\Delta(1-\Delta)=1+\nu^2,
\ee
which coincides with the (mass)$^2$ of a scalar field in $\dS_3$. 

We now briefly mention three other applications:  the relation with the split representation, the HKLL formula, and with an elegant  derivation of a fixed point formula for the dS$_3$ Green function. More details can be found in Appendix \ref{app:group}.

\medskip

\subsection*{Split representation}
\vspace{-.5mm}
The bulk-to-boundary propagator \eqref{eq:K} can be expressed as a $SL(2,\mathbb{C})$ matrix element 
\bea\label{eq:K_as_matrix_element}
K_\Delta(X; z,\bar{z}) \is \langle z,\bar{z} | \pi_\Delta(g_X) | 0 
\rangle,
\label{kkernel}
\eea
where $|0\rangle$ is a reference state chosen at $z=0$. This statement is verified in Appendix \ref{app:blktobnd}.

The split representation  \eqref{eq:split} can then be given a group theoretic interpretation by noting that the Green function can be viewed as a bi-matrix element
\bea\label{eq:G_as_bi_matrix_element}
G_\Delta(X, Y) = \langle 0 | \tilde\pi_{\Delta}(g_X^{-1}) \, \mathcal{S}_2\, \pi_{\Delta}(g_Y) | 0 \rangle_{\strut},
\eea
where $\mathcal{S}_2$ is an intertwining operator between the dual $SL(2,\mathbb{C})$ representations  $\pi_{\Delta}$ and $\tilde\pi_\Delta \equiv \pi_{1-\Delta}$. In practice, $\mathcal{S}_2$ amounts to performing the 2D shadow transform \cite{Simmons_Duffin_2014,Sun:2021thf}
\bea
  \label{shadowtransform}   
   {\cal S}_2 \mathbb{O}_\Delta(z,\bar{z}) =   
     \int_{\mathbb{C}} 
     \! d^2z' \,\frac{\mathbb{O}_{\Delta}(z,\bar z)}{|z-z'|^{4(1-\Delta)}\!\!\!}\,\; ,
\eea
which behaves as a 2D CFT operator with left and right scaling dimension $(1-\Delta,1-\Delta)$. 

 For the principal series, the operator $\mathbb{O}_{\Delta}$ and its shadow $\mathcal{S}_2\mathbb{O}_{\Delta}$ are related by complex conjugation,
rendering the matrix element \eqref{eq:G_as_bi_matrix_element}
invariant under the right action of the stabilizer $SU(1,1)$.
The split representation \eqref{eq:split} then follows  by using completeness 
\bea\label{eq:split_rep}
G_\Delta(X, Y) = \int_{{\mathbb C}} d^2 z \, \langle 0 | \tilde{\pi}_{\Delta}(g_X^{-1}) \mathcal{S}_2 | z,\bar{z}\rangle \langle z,\bar{z} | \pi_{\Delta}(g_Y) | 0 \rangle,
\eea
and  expressing the two matrix elements in \eqref{eq:split_rep} in terms of boundary-to-bulk propagators via equation \eqref{eq:K_as_matrix_element}. 

\medskip

 \begin{figure}[t]
\begin{center}
  \includegraphics[width=9.5cm]{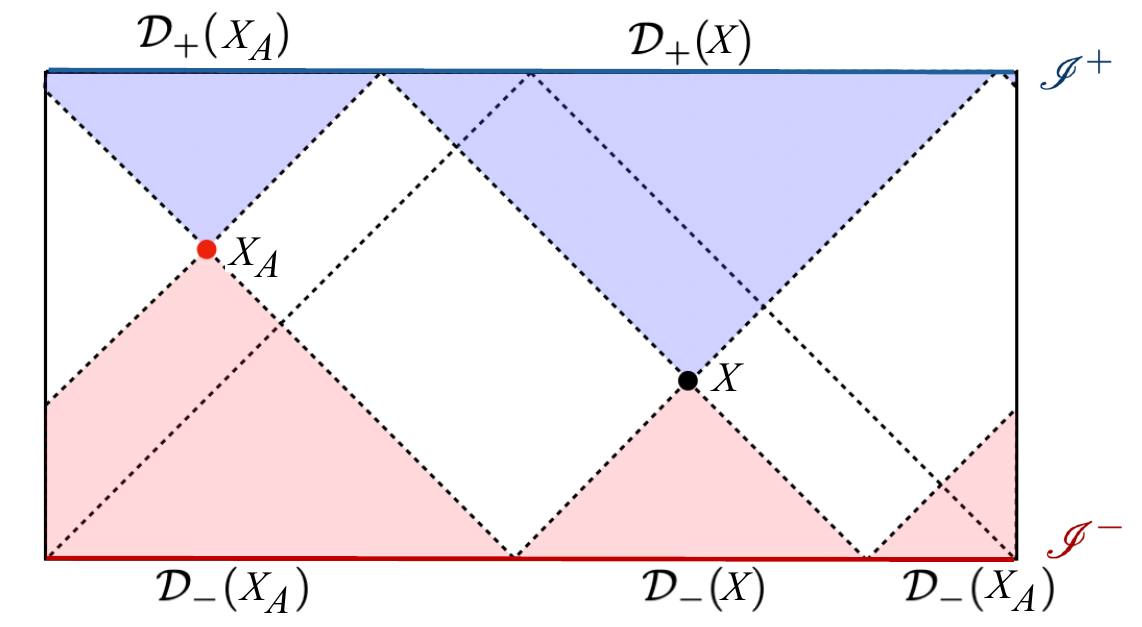}
  \caption{The physical operator $\mathbb{O}_\Delta(X)$ defined in \eqref{eq:OphysSU11} can be re-expressed as an HKLL integral \eqref{eq:HKLLbilocal} over all of ${\mathscr{I}^+}$. The HKLL inegral splits up into an integral over the future domain ${\cal D}_+(X)$ of the bulk point $X$ and over the future domain ${\cal D}_+(X_A)$ of the anti-podal point $X_A$. \label{deSitterLightCones}}
 \end{center}
 \vspace{-1mm}
\end{figure}

\subsection*{HKLL representation}
\vspace{-.5mm}
Here we show that our definition of the physical bulk operator coincides with the more standard dS$_3$/CFT$_2$ construction of bulk free fields in de Sitter spacetime.  

Substituting \eqref{eq:alphabeta}  into \eqref{eq:OphysSU11}
and inserting the definition \eqref{shadowtransform} of the shadow transform 
exhibits the physical operator as a convolution of bilocal CFT$_1$ operator 
with a kernel of the familiar form
\begin{eqnarray}
 \mathbb{O}_\Delta(X) \, =\, \int d^2z\,  \bigl\langle \spc 0 \spc \bigl| \pi_\Delta(g) \bigr| z,\bar{z}\bigr\rangle\bigl\langle z,\bar{z}\, \bigl| \mathbb{O}_\Delta \bigr\rangle\qquad
 \notag\\[-2mm]\label{eq:HKLLbilocal}
 \\[-2mm]\notag
\qquad \bigl\langle \spc 0 \spc \bigl| \pi_\Delta(g) \bigr| z,\bar{z}\bigr\rangle  \,=\, \l( \frac{\eta}{-\eta^2+(\bar{z}-\bar{x})(z-x)} \r)^{2(1-\Delta)} 
\end{eqnarray}
This formula looks similar to the HKLL formula \cite{Hamilton:2006az} in AdS$_3$/CFT$_2$, but there are some key differences \cite{Xiao:2014uea}\cite{Doi:2024nty}. Most importantly, rather than being restricted to a causal diamond, the 2D integral over $z$ and $\bar{z}$ runs over all of $\mathscr{I}^+$ (or $\mathscr{I}^-$) boundary, as depicted in figure \ref{deSitterLightCones}.

As shown, for a given bulk point $X$, future infinity $\mathscr{I}^+$ splits up into two parts, the causal future ${\cal D}_+(X)$ of $X$ and the causal future ${\cal D}_+(X_A)$ of its anti-pode. In AdS-holography, one would restrict the HKLL integral kernel to just the region in causal contact with the bulk point. Here we have an additional contribution from the integral over ${\cal D}_+(X_A)$, or equivalently, over the region ${\cal D}_-(X)$ of $\mathscr{I}^-$ in the causal past of $X$. This second integral gives the shadow contribution to the bulk operator $\mathbb{O}(X)$ and ensures that its Green function has the correct support on positive and negative frequency modes. For a more detailed discussion of the HKLL prescription for 3D de Sitter holography, we refer to {Appendix \ref{app:HKLL} and} \cite{Doi:2024nty}.

\subsection*{A fixed point formula}
\vspace{-.5mm}

The group theoretic perspective allows for an elegant derivation of the dS$_3$ Green function, as follows.
We  write the two-point function between operators located at $X$ and $Y$ as
\bea
& & \ \ G(X,Y) = \bigl\langle \mathbb{O}_\Delta(g) \mathbb{O}_\Delta(e) \bigr\rangle, 
\label{gphystwo} \notag \\[-2mm]\\[-2mm]\notag & & \mathbb{O}_\Delta(g)=
\int \! \dd z\, 
\widetilde{\mathcal{O}}^{\,L}_{1-\Delta}(z)\,  g\!\cdot
\mathcal{O}^{R}_\Delta(z)
\eea
where $e$ is the identity group element and $g$ is the $SL(2,\mathbb{C})$ isometry transformation that maps $X$ to $Y$. 
Without loss of generality, we can set 
\bea
    \label{gsigma}
a=d = \cosh\tau \qquad b =c = \sinh\tau,
\eea 
with $\tau$ the timelike geodesic distance from $X$ to $Y$. Below we denote 
\bea
g(z)= \frac{az+b}{cz+d}
\eea
and write $ G(X,Y) = G(\tau)$.

To evaluate the overlap integral in the two-point function, we recall the shadow pairing identity given in equation (3.51) of ~\cite{Sun:2021thf}.
Normally, this shadow identity is written with a principal value prescription; we will instead use a Feynman contour.
Employing the resulting identity trivializes the evaluation of the physical two-point function \eqref{gphystwo} to
\bea\label{fixedpoint}
G_F(\tau) \is \theta(\tau) G^{+}(\tau) + \theta(-\tau) G^{-}(\tau)\notag\\[-1.5mm]\\[-1.5mm]\notag
G^{\pm}(\tau) \is  \int\! \frac{dz}{2\pi i} \, \frac{g'(z)^{\Delta}}{z\nspc-\nspc g(z)\nspc\mp\nspc i \epsilon\!}
\eea
Applying the residue formula produces a fixed point formula that expresses the Green function as a sum over classical paths.
The eigenvalues of the $SL(2,\mathbb{C}$) matrix $g$ given in \eqref{gsigma} are $
\lambda_\pm=e^{\pm\tau}$ and the multipliers at the two solutions to the fixed point equation $z = g(z)$ are 
\bea
g'(z_\pm)=\lambda_\mp^2= e^{\mp 2\tau}.
\eea
Putting everything together we arrive at  
\begin{equation}
G_F(\tau)
=
\frac{\sinh\bigl((2\Delta-1)(\tau - i \pi )\bigr)}{\sinh\bigl(\tau - i\epsilon\, {\rm sgn}(\tau)\bigr)}.
\label{eq:negative_result}
\end{equation}
which (up to an overall normalization) equals the known Feynman Green function in 3D de Sitter as a function of the geodesic distance~$\tau$, \cite{Bousso:2001mw}. For more details about the intermediate steps in the above derivation, see Appendix \ref{app:fixed-point}.

\medskip

\section{Microscopic vs coarse-grained equal-energy constraint}
\label{sec:fine-coarse}
\vspace{-.5mm}

In our discussion thus far we have ignored an important subtlety of our construction of physical operators. As mentioned in the introduction, one can view the equal energy constraint \eqref{physconstraint} as either a microscopic condition that amounts to a one-to-one pairing between energy eigenstates or as a coarse grained condition that the two energies match within a very small energy window and that only becomes exact in the large
$N$ limit. The two definitions of the equal energy condition lead to different physical operators and answers for the physical two-point functions. Our results described thus far apply to the coarse grained definition.

To explain the difference, imagine we pick a particular microscopic realization of the SYK model with a discrete energy spectrum.
How does the large $N$ limit work for the doubled SYK system coupled via the equal energy constraint? 

For the exact microscopic pairing  $|E_i\rangle = |E_i\rangle_L |E_i\rangle_R$ between energy states, the spectral density and closure relation for the coupled SYK system would be the same as for a single SYK system.
Upon taking the large $N$ limit, the spectrum becomes continuous. The discrete closure relation then gets replaced by an integral 
\bea
\mathds{1} = \sum_{i}\; |E_i\rangle \langle E_i | \ \to\  \mathds{1}_{\rm fine} = \int\! dE\, \rho(E) \, |E\rangle \langle E |  
\eea
We call this the fine-grained closure relation.

{Alternatively, we can take the large $N$ limit first. Exponentially close energy levels coalesce into a continuous energy density $\rho(E)$ and the resolution of identity in the doubled theory can be written as:
\bea
\mathds{1} = \int dE_L dE_R \ \rho(E_L) \rho(E_R) | E_L \rangle | E_R \rangle \langle E_L | \langle E_R |. 
\eea
The relative time integral in the definition of physical operators (\ref{physoperator}) effectively introduces a factor $
\int_{-\infty}^{+\infty}\! dt \, e^{-it (E_L-E_R)} = \delta(E_L-E_R) $.
This leads to the closure relation
}
\bea
\mathds{1}_{\rm coarse} = \int\! dE \, \rho(E)^2 |E\rangle \langle E|
\eea
We will call this the coarse-grained closure relation and energy spectrum \footnote{{Physically, what happens is that once we have $N=\infty$, the limit over $T$ in $\lim_{T\to \infty}\frac{1}{\sqrt{2T}} \int_{-T}^{T} dt$ cannot go all the away to infinity. This is why this integral identifies energy levels up to a small tolerance, leading to $\rho(E)^2$. Depending on the microscopic details of this limit we can, in principle, get any power $\rho(E)^a, \ a \in [1,2]$. }}.

A simple calculation shows that the two-point function of physical operators for the two prescriptions are given~by
\beq
\label{eq:fine_rho}
\begin{split}
\langle \mathbb{O}(t)\, \mathbb{O}(0) \rangle_{E_0}
&= \int dE\, \rho(E)^a\, e^{-i t (E - E_0)} 
\big| \langle E_0| {\cal O}_L |E\rangle \big|^2 
\big| \langle E_0 | {\cal O}_R |E\rangle \big|^2\\[2mm]
&\qquad  a=\biggl\{\  \begin{array}{cc}  {1 \ \ \ } &  {\ \ \mbox{fine-grained} }
\\[.5mm]  {2 \ \ \ } & {\ \, \mbox{coarse-grained}}
\end{array}
\end{split}
\eeq
It turns out that both definitions of physical operators preserve our main result that the physical two point functions match with the de Sitter Green function. Indeed, it is easy to see that the spectral density pre-factor $\rho(E)^a$ can be absorbed in a simple time-shift:
writing $E = E_0 + \omega$ and using $\rho(E) = \rho(E_0)e^{\beta \omega}$, we have 
\bea
\rho(E)^a e^{-it(E-E_0)} = \rho(E_0)^a e^{-i(t+ia \beta)\omega}.
\eea

\begin{table}[t]
\centering 
\begin{tabular}{|c|c|c|}
\hline
    \ \ 1D model \ \ & 
     coarse-graining & Green function \\ 
     \hline
     \hline
     \hline
    \ \ Schw QM\  \  & \;  fine~~~\,$a = 1$ & $G_{\rm dS}(\tau - i \pi)$  \\
     \hline
    \ \ Schw QM  \Large\strut\  & coarse\; $a = 2$ & $G_{\rm dS}(\tau)$  \\   
     \hline
    \ \ DSSYK$_T$ \Large\strut \  & \;  fine~~~\,$a=1$ & $G_{\rm dS}(\tau-i \pi)$ \\
     \hline
     \ \; DSSYK$_T$  \Large\strut \ & coarse\; $a = 2$ & \ \ $G_{\rm dS}(\tau - i \pi (1-v))$ \\ 
     \hline
    \ \ \ DSSYK$_{\infty}$  \Large\strut \ & $a\in[1,2]$ & $G_{\rm dS}(\tau-i \pi)$ \\    
    \hline
\end{tabular}
 \caption{Table listing the match between the two-point functions $\langle \mathbb{O}(t) \mathbb{O}(0)\rangle$ and de Sitter Green function for Schwarzian QM and DSSYK   (at arbitrary temperature) with different energy constraints. Here DSSYK means the leading $p^2/N \rightarrow 0$ answer, that is, large $p$ SYK at infinite $N$.
  Here $v$ and $\beta=1/T$ are related via $\beta = \frac{\pi v}{\cos\frac{\pi v}2}$ and geodesic distance $\tau$ is related to the SYK time $t$ through the ``fake" temperature:
 $\tau = \frac{\pi v t}{\beta}$.
For Schwarzian $v=1$, since it corresponds to the low-energy limit of SYK. For DSSYK at infinite temperature, the two-point function does not depend on $a$. 
 }
\label{tab:results}
\end{table}

Table \ref{tab:results} lists how the choice of $a$ affects the match between the two-point function and the de Sitter Green function for Schwarzian quantum mechanics and for the DSSYK model at finite and infinite temperature. For a discussion of the Schwarzian QM with an equal energy constraint, see Appendix \ref{sec:appendix}; the DSSYK analysis is carried out in Appendix \ref{app:dssyk}.
We see that for DSSYK at infinite temperature ($v=0$), the coarse-graining does not affect the physical Green function, while the fine-grained prescription $a=1$ for any temperature always leads to the anti-podal de Sitter Green function \cite{narovlansky2023doublescaledsyksitterholography,Narovlansky:2025tpb}. For generic $a$, in the Schwarzian case one obtains $G_{dS}(\tau + i \pi (a-2))$ while for DSSYK it is $G_{dS}(\tau - i \pi + i \pi v(a-1))$.

\medskip

\def\lL{{\mbox{\tiny $L$}}}
\def\rR{{\mbox{\tiny $R$}}}

\section{Concluding remarks}
\label{sec:conclusion}
\vspace{-1mm}

We have shown that generic large $N$ 1D CFTs contain an algebra of generalized free fields (GFF) with the same correlation functions as local QFT operators on the geodesic worldline of an idealized observer in d+1-dimensional de Sitter space time. This known observation follows from the special symmetry properties of de Sitter spacetimes \cite{Anninos_2012} and the split representation of de Sitter Green functions \cite{Sleight:2019mgd}. We outlined an explicit construction in the double scaled SYK model of how to implement a large $N$ factorization mechanism that ensures that the generalized free fields satisfy Wick's rule.

The GFF construction is a small but necessary first step towards a more complete theory of worldline de Sitter holography. Our construction shares some elements of dS/CFT but differs from it in important ways. In standard holography, there is only one extra holographic direction. By restricting to worldline observables, we were able to construct operators in a 1D quantum system that contains generalized free fields in a dual dS spacetime with an (a priori) arbitrary number of extra dimensions. 

Worldline holography has the advantage that time evolution in de Sitter and the dual quantum theory are identified \cite{Anninos_2012,Chandrasekaran:2022cip,narovlansky2023doublescaledsyksitterholography}. This allows for a direct comparison between the physical properties  and observables on both sides. A  disadvantage, however, is that the worldline is embedded in the ambient spacetime and therefore subject to the rules of de Sitter quantum gravity. The holographic dictionary can therefore not be as cleanly formulated as in anti-de Sitter space. A minimal subset of properties that the holographic model of de Sitter should incorporate are \cite{narovlansky2023doublescaledsyksitterholography}
(i) a dimensionless coupling that controls the size of $G_N/R_{\rm dS}$, (ii)~the existence of a generalized free field algebra in the $G_N\to 0$ limit, (iii) a bounded energy spectrum and finite time resolution, (iv)
de Sitter thermodynamics and quasi-normal modes (including the cosmological horizon superdiffusion \cite{Milekhin:2024vbb}), (v) gravitational interactions and backreaction.
The above results and evident properties of DSSYK imply that it realizes the first three properties. 
Recent work \cite{verlinde2024doublescaledsykchordssitter,Marini:2026zjk} indicates that DSSYK also captures essential quantitative features of gravitational backreaction and thermodynamics of de Sitter gravity via the identification $\lambda  = 4 \pi G_N$. 

There are many concrete future directions to explore. One logical next step is to study the correlators of physical operators as a function of SYK temperature, and compare the result with the behavior of the Green function in a Schwarzschild-de Sitter background. Another interesting question is to study how the known formulas for the OTOC in SYK  \cite{Berkooz:2018jqr} and its quantum group origin~\cite{Belaey:2025ijg,Xu:2025zkr,Schouten:2025tvn}  can be incorporated into the above holographic dictionary and compared with the prediction from shockwave interactions in de Sitter \cite{Aalsma:2020aib,Narovlansky:2025tpb}. Also it would be interesting to understand the computation of time-like entanglement \cite{Doi:2023zaf,Milekhin:2025ycm} entropy in this model and how it maps to geodesics in de Sitter.

\medskip

\section*{\bf Acknowledgements}

\vspace{-1mm}

We thank Dionysios Anninos, Andreas Blommaert, David Kolchmeyer, Juan Maldacena, Tommaso Marini, Beatrix M\"uhlmann, Vladimir Narovlansky, Andrew Sontag, Damiano Tietto, Xiaoliang Qi, Zimo Sun, and Ying Zhao for helpful discussions and comments.
K.G. is supported by JSPS KAKENHI Grant-in-Aid for Early-Career Scientists (24K17048) and Research Fellowships of Japan Society for the Promotion
of Science for Young Scientists (22J00663). AM acknowledges funding provided by the Simons Foundation (Grant 376205),  the DOE QuantISED program (DE-SC0018407), and the Air Force Office of Scientific Research (FA9550-19-1-0360). J.X. is supported by the U.S. Department of Energy, Office of Science, Office of High Energy Physics, under Award Number DE-SC0011702. J.X. acknowledges the support by the Graduate Division Dissertation Fellowship and the Physics Department Graduate Fellowship at UCSB.

\appendix

\section{Physical operators in covariant Schwarzian QM}
\label{sec:appendix}
\vspace{-1.5mm}

\noindent

In this Appendix we introduce a covariant version of Schwarzian quantum mechanics which naturally contains physical operators of the form \eqref{physoperator} with correlation functions that reproduce those of a GFF in de Sitter spacetime. Consider the following generalized Schwarzian quantum mechanics at finite temperature $\beta = 2\pi $:
\bea
\label{schwaction}
S[f,g]\, 
\is  - m \int_0^{2\pi}\! dt \; \sqrt{{\rm Sch}( F,\spc t\spc)-{\rm Sch}(G,\spc t \spc)} ,\\[2mm]
 {\rm Sch}( \tau,\spc t\spc) &\! \equiv\! &  \frac {\dddot{\tau}}{\dot\tau} - \frac 3 2 \Bigl(\frac{\ddot\tau}{\dot\tau}\Bigr)^2, \, \qquad \quad\; \begin{array}{c} {F\, = \, \tan({f}/{2})},\\[1.5mm] {G\, = \, \tan({g}/{2})}.\end{array}
\eea
The functions $f(t)$ and $g(t)$ satisfy the periodicity condition $f(t+2\pi) = f(t) +2\pi$ and $g(t+2\pi) = g(t) +2\pi$  and define elements of the group Diff$(S^1)$ of diffeomorphisms of the thermal circle.  
The covariant Schwarzian action \eqref{schwaction}  can be cast in a first order form \cite{Engels_y_2016}
\bea
S \! \is \!\! \int \! dt \,\Bigl(p_{\varphi}\dot \varphi +  p_{\rho} \dot \rho + \lambda \dot{F} +  \mu \dot{G }  - e (H_L - H_R + m^2)) \Bigr)\notag\\[-2mm]\\[-2mm]\notag
& & \qquad \ H_L\! =  p_{\varphi}^2 + \lambda e^\varphi, \ \ \qquad \  \ 
H_R = \,   p_{\rho}^2 + \mu e^\rho.
\eea
As shown in \cite{Engels_y_2016}, this Hamiltonian is on-shell equal to the Schwarzian and equal to the SL(2,R) Casimir.
The variable $e$ transforms like a one-form and acts like a Lagrange multiplier imposing the energy constraint 
\bea
H_\lL - H_\rR + m^2=0. 
\eea
Alternatively, we can view the variables $\lambda$ and $\mu$ as Lagrange multipliers that impose 
\bea
\label{constrfg}
e^{-1}\dot{F} = e^{\varphi}, \ \ \qquad \ \ e^{-1} \dot{G} = e^\rho. 
\eea
From here on we will take the `massless' limit $m^2=0$. 
The model then reduces to two Schwarzian QM models coupled via a zero energy constraint.

The above action is invariant under reparametrizations of $t$.
After integrating out the momentum variables, solving for the constraints  \eqref{constrfg} imposed by $\lambda$ and $\mu$, and introducing the dynamical time-coordinate $\tau$ via $e = \dot{\tau}$, the covariant Schwarzian action attains the form 
\bea
\qquad S[f,g,\tau] \is 
\int\! dt \,  {\dot{\tau}}\spc  \Bigl( {\rm Sch}(F,\tau) - {\rm Sch}(G,\tau)\Bigr)\notag
\\[-2mm]& & \qquad \hspace{6cm} \quad  e \equiv \dot\tau\\[-2mm]
\notag
\is \int\! dt \,  \frac{1}{\dot{\tau}}\spc  \Bigl( {\rm Sch}(F,t) - {\rm Sch}(G,t)\Bigr)
\eea
This action is reparametrization invariant since $\tau$ is dynamical. 

Fixing the gauge $\tau = t$, the action simplifies to the sum of two ungauged Schwarzian actions
with the physical constraint imposed that the difference of the two Hamiltonians must vanish. Due to this constraint, the spectrum of the covariant SQM is identical to that of a single Schwarzian QM model. The new feature of the coupled theory is that its set of physical observables is different.

The discussion of the physical operators then proceeds as in the main text.  As before, the physical operators that preserve the energy constraint take the form of integrals over time of the product of a left and right operator with conformal dimension $\Delta$ and $1-\Delta$. 
\bea
\label{physoperatortwo}
\mathbb{O}_\Delta(\tau) = {\cal N} \int\! dt\; {\cal O}^L_{1-\Delta}(t) {\cal O}^R_{\Delta}(\tau-t)
\eea
Let us compute the two point function via the spectral representation.  This gives for the two-point function, defined as the expectation value in the energy eigenstate $|E_0\rangle$ 
\bea
\bigl\langle \mathbb{O}_\Delta(\tau)\mathbb{O}_\Delta(0)\bigr\rangle
 \hspace{-4mm} & & = \int \! dE \, \widetilde{G}(E)\, e^{-(E-E_0)\tau}\notag\\[-2mm]\\[-2mm]\notag
\widetilde{G}(E) \is \rho(E) \bigl|\la E  \bigl| {\mathbb{O}}_{\!\Delta} \bigl| E_0\ra \bigr|^2
\eea
Here, following \cite{narovlansky2023doublescaledsyksitterholography}, we used the fine-grained energy constraint.

The matrix elements between energy eigenstates in SQM are exactly known \cite{Mertens:2017mtv}. Introducing $s =\sqrt{E}$ and $s_0 = \sqrt{E_0}$ 
\bea
\label{matrixeltsqm}
\widetilde{G}(E)  \is b_\Delta  \,
\frac{\Gamma(\Delta\pm is\pm is_0)\,{\Gamma(1\!-\!\Delta\pm is\pm is_0)}}{{\Gamma(2i s)\Gamma(1\!-\!2i s) \Gamma(2i s_0) \Gamma(1\!-\!2i s_0)
}{}}
\nonumber \\[-1mm]\\[-1mm]\nonumber
\is \; b_\Delta \, \frac{\sinh(2\pi s)\sinh(2\pi s_0)}{\sin\bigl(\pi (\Delta\pm is\pm is_0)\bigr)\!\!\!}
\eea
with $b_\Delta = 2\Delta \sinh 2\Delta$. Here we included a factor of $\rho(E_0)$ for normalization. Let $s_0 = \sqrt{E_0} = \pi/\beta$ and writing $s = s_0+  \omega$, we have
\bea
\,\widetilde{G}(E)
\   \simeq \,\frac{b_\Delta}{\sin\bigl(\pi (\Delta\pm i \omega)\bigr)\!\!\!}
\label{gammax}
\eea
Fourier transforming and normalizing gives the anti-podal 3D de Sitter Green function 
\begin{equation}
G_{\rm dS,A}(\tau)
\, = \, \;
\frac{\sinh\bigl((2\Delta-1)\tau \bigr)}{\sinh\bigl(\tau - i\epsilon\,\bigr)}.
\end{equation}
This is the result is listed at the top row in table \ref{tab:results}.

Next let us compute the two-point function with the coarse grained energy constraint. As explained in section 6, the spectral Green function $\widehat{G}(E)$ then takes the form
\bea
\widehat{G}(E) \is \rho(E)^2 \bigl|\la E  \bigl| {\mathbb{O}}_{\!\Delta} \bigl| E_0\ra \bigr|^2\notag \\[-2mm]\\[-2mm]\notag
 \is \; b_\Delta {\frac{\, \sinh(2\pi s)^2}{\sin\bigl(\pi (\Delta\pm is\pm is_0)\bigr)\!\!\!}}\;
\; \simeq   \; {\frac{b_\Delta e^{2\pi\omega}}{\sin\bigl(\pi (\Delta\pm i \omega)\bigr)\!\!\!}}
\eea
Fourier transforming  gives the same-sided Green function in 3D de Sitter spacetime
\begin{equation}
G_{\rm dS}(\tau)
=
\frac{\sinh\bigl((2\Delta-1)(\tau - i\pi) \bigr)}{\sinh\bigl(\tau - i\epsilon\, \bigr)}.
\label{eq:negative_result}
\end{equation}
This is the result listed on the second row in table \ref{tab:results}.

\section{Group theory interpretation of the split representation
}
\label{app:group}
In this section, we collect some details on the split representation and its group theory interpretation. We first show that the bulk-to-boundary propagator can be written as a matrix element of an $SL(2,\mathbb C)$ representation. We then derive the split representation of the dS$_3$ bulk-to-bulk Green function. Finally, we explain how the shadow transform converts the bulk-to-boundary propagator of weight $\Delta$ into that of weight $1-\Delta$, thereby establishing the matrix-element form of the split representation. Useful background references are \cite{harish1969harmonic,vilenkin1978special,Sun:2021thf}

\subsection{Bulk-to-boundary propagator as an $SL(2,\mathbb{C})$ matrix element}
\label{app:blktobnd}

For $d=2$, the bulk-to-boundary propagator takes the form
\begin{equation}
K_{\Delta}(X ; z, \bar{z})=c_{\Delta} e^{-i 2 \pi \Delta}\left(\frac{\eta}{\eta^2-(z-x)(\bar{z}-\bar{x})}\right)^{2 \Delta},
\end{equation}
where we use the notation of equation~(16) and identify the two-dimensional Euclidean plane with $\mathbb C$, so that $(z,\bar z)$ and $(x,\bar x)$ denote boundary and bulk spatial coordinates, respectively. We now show that this propagator can be written as a matrix element of $SL(2,\mathbb C)$.

$SL(2,\mathbb{C})$ is the global subgroup of the Virasoro conformal group in two dimensions. Consider the standard action of $SL(2,\mathbb C)$ on a scaling operator $f(z,\bar{z})$
\begin{equation}
\left[\pi_{\Delta}(g) f\right](z, \bar{z})
=
\bigl(c z+d\bigr)^{-2 \Delta}\left(\bar c  \bar{z}+\bar d \right)^{-2 \Delta}
f\left(\frac{a z+b}{c z+d}, \frac{\bar a \bar{z}+\bar b}{\bar c \bar{z}+\bar d}\right) .
\end{equation}
The relevant conformal weight is $2\Delta$. Here we identify $g$ with points on 3D de Sitter spacetime via equation 
\eqref{eq:alphabeta}. Equivalently, we pick the reference bulk point to be
$
X_0
= 
(1,0,0)
$
and define the reference state $|0\rangle$ by
\begin{equation}
\langle z, \bar{z}| 0\rangle
=
K_{\Delta}\left(X_0 ; z, \bar{z}\right)
=
\frac{c_{\Delta} e^{-i 2 \pi \Delta}}{(1-z \bar{z})^{2 \Delta}}.
\end{equation}
We claim that for a general bulk point $X=(\eta,x,\bar{x})$, the bulk-to-boundary propagator is
\be
K_\Delta(X;z,\bar{z})= \langle z, \bar{z}| \pi_{\Delta}\left(g_X\right)|0\rangle.
\ee
To verify this, note that
\be \label{eq:id-gz}
 \langle z, \bar{z}| \pi_{\Delta}\left(g_X\right)|0\rangle
=
\bigl(c z+d\bigr)^{-2 \Delta}\left(\bar c \bar{z}+\bar d\right)^{-2 \Delta}
\frac{c_{\Delta} e^{-i 2 \pi \Delta}}
{\bigl(1- (g_X \cdot z) (\overline{g_X \cdot z{\mbox{\scriptsize ${\strut}$}}})\, \bigr)^{2 \Delta}},
\ee
where we choose
\be
g_X=
\biggl(\begin{array}{cc}
a  & b \\
c &  d
\end{array}\biggr)
\equiv
\left(\begin{array}{cc}
\eta^{-1 / 2} & -x \eta^{-1 / 2} \\
0 & \eta^{1 / 2}
\end{array}\right)\quad \rightarrow \quad 
\begin{array}{c}{g_X \cdot z=(z-x)/\eta}\\[2.5mm]
{\overline{g_X \cdot z{\mbox{\scriptsize ${\strut}$}}}=(\bar{z}-\bar{x})/\eta}\end{array},
\ee
and hence
\be
1-\left(g_X \cdot z\right)\overline{\left(g_X \cdot z\right)}
=
1-\frac{(z-x)(\bar{z}-\bar{x})}{\eta^2}
=
\frac{\eta^2-(z-x)(\bar{z}-\bar{x})}{\eta^2}.
\ee
Substituting this into~\eqref{eq:id-gz}, we obtain
\be
\langle z, \bar{z}| \pi_{\Delta}\left(g_X\right)|0\rangle
=
c_{\Delta} e^{-i 2 \pi \Delta}
\left(\frac{\eta}{\eta^2-(z-x)(\bar{z}-\bar{x})}\right)^{2 \Delta},
\ee
which is the complexified bulk-to-boundary propagator $K_\Delta$.

\subsection{Split representation of the Green function on dS$_3$}
\label{app:splitg}
For a general configuration
\be
X=(\eta, x,\bar{x}),\qquad Y=(\eta', y,\bar{y}),\qquad \rho^2:=| x- y|^2,
\ee
we derive the following split representation of the dS$_3$ bulk-to-bulk Green function:
\bea\label{eq:G-split2}
G_{\Delta}^{\mathrm{dS}}(X, Y)= \frac{b}{2\pi} \int_{\mathbb{R}^2} d^2 z\left(\frac{\eta}{\eta^2-|{z}-{x}|^2+i \epsilon}\right)^{2 \Delta}\left(\frac{\eta^{\prime}}{\eta^{\prime 2}-|{z}-{y}|^2+i\epsilon}\right)^{2(1-\Delta)},
\eea
where $b = 2-2\Delta$ is an overall normalization constant.

Using Feynman parametrization, the integral can be written as
\bea
G^{\rm dS}_\Delta(X,Y)
\is
\frac{b\,\eta^a\eta'^b}{2\pi \Gamma(a)\Gamma(b)}
\int_0^1 du\,u^{a-1}(1-u)^{b-1}
\int_{\mathbb R^2} d^2 z\;
\frac{1}{\bigl[uA+(1-u)B\bigr]^2},
\label{eq:after-Feynman}\\[2mm]
a=2\Delta,\qquad b\is2-2\Delta,\qquad
A=\eta^2-| z- x|^2+i\epsilon,\qquad
B=\eta'^2-| z- y|^2+i\epsilon.
\ee
The denominator can be reorganized as
\bea
& & \ \ u A+(1-u)B= M(u)-| z- z_u|^2+i\epsilon,
\label{eq:Mu-def}\\[2mm]
M(u)\is u\eta^2+(1-u)\eta'^2-u(1-u)\rho^2, \qquad {z}_u=u {x}+(1-u) {y}.
\eea
The $z$-integral can then be performed explicitly:
\be
\int_{\mathbb R^2} \frac{d^2 z}{\bigl(M(u)-| z -  z_u|^2+i\epsilon\bigr)^2}
=
-\frac{\pi}{M(u)+i\epsilon}.
\label{eq:z-integral}
\ee
Substituting this into~\eqref{eq:after-Feynman}, we obtain
\be \label{eq:G2-def}
G_{\Delta}^{\mathrm{dS}}(X, Y)= - \frac{b \spc \eta^a \eta'^b}{2\spc  \Gamma(a) \Gamma(b)} \int_0^1 d u \frac{ u^{a-1}(1-u)^{b-1} }{M(u) +i\epsilon}.
\ee

Next we write
\be
\rho^2=\eta^2+\eta'^2+ 2\eta\eta' Z, \qquad \quad Z :=\frac{-\eta^2-\eta'^2+| x- y|^2}{2\eta\eta'}.
\label{eq:rhoZ}
\ee
and make the change of variables
\begin{equation}
u=\frac{\eta' t}{\eta+\eta' t},
\qquad
1-u=\frac{\eta}{\eta+\eta' t},
\qquad
du=\frac{\eta\eta'}{(\eta+\eta' t)^2}\,dt.
\label{eq:tsub}
\end{equation}
Then~\eqref{eq:G2-def} becomes
\begin{equation}
G^{\rm dS}_\Delta(X,Y)
=
\frac{b}{2\spc \Gamma(a)\Gamma(b)}
\int_0^\infty\!\!\! dt\;
\frac{t^{a-1}}{t^2 - 2Z t+1-i\epsilon},
\label{eq:t-integral}
\end{equation}
where all factors of $\eta$ and $\eta'$ cancel because $a+b=2$. The remaining integral evaluates to the familiar hypergeometric form
\be
\int_0^\infty dt\,
\frac{t^{a-1}}{t^2-2Z t+1-i \epsilon}
=
\frac{\pi(1-a)}{\sin (\pi a)}
\,{}_2 F_1\left(a, 2-a ; \frac{3}{2} ; \frac{1- Z+i \epsilon}{2}\right).
\ee
Moreover, this particular ${}_2F_1$ admits the simpler expression
\be
{}_2 F_1\left(a, 2-a ; \frac{3}{2} ; \frac{1-Z}{2}\right)
\, =\,
\frac{\sin ((a-1)\arccos Z)}{(a-1)\sqrt{1-Z^2}}.
\ee
Therefore, we find~\eqref{eq:G-split2} produces the dS$_3$ two-point function
\be
\begin{split}
G_{\Delta}^{\mathrm{dS}}(X, Y)
&\, =\; 
\frac{\sin((1-2\Delta )\arccos Z)}{\sqrt{1-Z^2}} .
\end{split}
\ee

\subsection{Shadow transform of the bulk-to-boundary propagator}

In this section, we study the shadow transform of the bulk-to-boundary propagator by analyzing the action of $\mathcal S_2$ on the corresponding matrix element. Our goal is to show that it produces the propagator of complementary weight:
\be
\la0|\tilde{\pi}_{\Del}(g_{X})\mathcal{S}_2|z,\bar{z}\ra \, \propto\, K_{1-\Delta}(X;z,\bar{z}).
\ee
Inserting a complete basis, we can write the matrix element as
\be \label{eq:S-int}
\la0|\tilde{\pi}_{\Del}(g_{X})\mathcal{S}_2|z,\bar{z}\ra
=
\int_{\mathbb{C}}\frac{\dd^{2}w}{2\pi i}\,
\la0|\tilde{\pi}_{\Del}(g_{X})|w,\bar{w}\ra\,
\la w,\bar{w}|\mathcal{S}_2|z,\bar{z}\ra
=
\int_{\mathbb{C}}\frac{\dd^{2}w}{2\pi i}\,
\frac{K_{\Del}(X;w,\bar{w})}{|z-w|^{4(1-\Del)}}.
\ee
Here we used the fact that the matrix element of the dual representation reproduces the bulk-to-boundary propagator,
\be
\langle 0| \tilde{\pi}_{\Delta}\left(g_X\right)|w, \bar{w}\rangle
=
\langle w, \bar{w} | \pi_\Delta (g_X) |0\rangle
=
K_\Delta(X;w,\bar{w}),
\ee
while the complexified shadow transform gives the kernel
\be
\langle w,\bar{w} | \mathcal{S}_2 | z,\bar{z}\rangle
=
\frac{1}{|z-w|^{4(1-\Delta)}}.
\ee
To evaluate the integral in~\eqref{eq:S-int}, it is convenient to first consider its Euclidean version,
\be
K_{\Delta}^E(X ; w, \bar{w})
=
c_{\Delta} e^{-2 \pi i \Delta}\left(\frac{\eta}{\eta^2+|w-x|^2}\right)^{2 \Delta},
\ee
and then analytically continue the result back to the Feynman propagator. We therefore study
\bea\label{eq:S-int-E}
I_{\Delta}(X ; z, \bar{z})
:=
\int_{\mathbb{C}} \frac{d^2 w}{\pi}\,
\frac{K_{\Delta}^E(X ; w, \bar{w})}{|z-w|^{4(1-\Delta)}}
=
c_{\Delta} e^{-2 \pi i \Delta} \eta^{2 \Delta}
\int_{\mathbb{C}} \frac{d^2 y}{\pi}
\frac{1}{\left(\eta^2+|y|^2\right)^{2 \Delta} |\xi-y|^{4(1-\Delta)} }.
\eea
In the second expression above, we introduced $
y=w-x$ and $\xi=z-x$
for convenience. 
Our aim is to show that $I_{\Delta}(X;z,\bar{z})$ is proportional to $K^E_{1-\Delta}(X;z,\bar{z})$.

The integral~\eqref{eq:S-int-E} is analogous to a standard Euclidean loop integral, so one may apply Feynman parametrization and perform the $y$-integral first. This gives
\bea
I_{\Delta}(X ; z, \bar{z})
=
\frac{c_{\Delta} e^{-2 \pi i \Delta} \eta^{2 \Delta}}{\Gamma(2 \Delta) \Gamma(2-2 \Delta)}
\int_0^1 d u \,
\frac{u^{2 \Delta-2}(1-u)^{1-2 \Delta}}{\eta^2+(1-u)|z-x|^2}.
\eea
Introducing $v=1-u$, the $v$-integral is elementary. We find 
\be \label{eq:I-K-id}
I_{\Delta}(X ; z, \bar{z})
=
\frac{c_{\Delta} e^{-2 \pi i \Delta}}{2 \Delta-1}
\left(\frac{\eta}{\eta^2+|z-x|^2}\right)^{2(1-\Delta)}
=
\mathcal{N}_\Delta K^E_{1-\Delta}(X;z,\bar{z}),
\ee
with normalization constant
\be
\mathcal{N}_{\Delta}
=
\frac{c_{\Delta}e^{-2 \pi i \Delta}}{(2 \Delta-1)\, c_{1-\Delta} e^{-2 \pi i(1-\Delta)}}.
\ee
Finally, analytically continuing~\eqref{eq:I-K-id} via
\be
\eta^2 + |z-x|^2 \to \eta^2 - |z-x|^2 + i\epsilon,
\ee
we conclude that the shadow transform sends $K_\Delta$ to $K_{1-\Delta}$, as claimed.

\subsection{Relation to the fix-point formula}
\label{app:fixed-point}
From the split representation, we derived the dS$_3$ bulk-to-bulk Green function expressed with a one-dimensional integral (\ref{eq:t-integral}) .
Let us relate the one-dimensional integral to the
fixed point formula (\ref{fixedpoint}).  For timelike-separated points we write
$Z=-\cosh\tau$,
so that
\begin{equation}
G^{\rm dS}_\Delta(X,Y)
=
\int_0^\infty dt\,
\frac{t^{2\Delta-1}}
     {t^2+2\cosh\tau\, t+1-i\epsilon}.
\label{eq:t_integral_timelike}
\end{equation}
The hyperbolic Möbius transformation that appears in the fixed point
formula is
\begin{equation}
  g_\tau(z)
  =
  \frac{\cosh\tau\,z+\sinh\tau}
       {\sinh\tau\,z+\cosh\tau},
  \qquad
  g'_\tau(z)
  =
  \frac{1}{(\sinh\tau\,z+\cosh\tau)^2}.
\label{eq:gtau_fixed}
\end{equation}
We now introduce the change of variables
\begin{equation}
  t
  =
  -\frac{1}{\sinh\tau\,z+\cosh\tau}.
\label{eq:t_z_change}
\end{equation}
This gives
\bea
  t^2=g'_\tau(z), \ \qquad  \  dt
  =
  \sinh\tau\,t^2\,dz
\label{eq:t_gprime_dt}, \qquad \quad
\bigl(z-g_\tau(z)\bigr) \is -
  \frac{t^2+2\cosh\tau\,t+1}{\sinh\tau\,t\,}
\eea
Including the Feynman prescription, this means that near the relevant
poles
\begin{equation}
  t^2+2\cosh\tau\,t+1-i\epsilon
  =
  -\,\sinh\tau\,t\,
  \bigl(z-g_\tau(z)-i\epsilon\,{\rm sgn}(\tau)\bigr).
\label{eq:denominator_map_iepsilon}
\end{equation}
Here we used that the poles of the $t$-integral lie on the negative
real $t$-axis, so that
${\rm sgn}(\sinh\tau\, t)=-{\rm sgn}(\tau)$
at the poles.  Combining the above identities gives
\begin{equation}
  \frac{dt\,t^{2\Delta-1}}
       {t^2+2\cosh\tau\,t+1-i\epsilon}
  =
  -\,
  \frac{dz\,g'_\tau(z)^\Delta}
       {z-g_\tau(z)-i\epsilon\,{\rm sgn}(\tau)}.
\label{eq:t_integrand_to_z_integrand}
\end{equation}
The overall minus sign can be absorbed into the orientation of the image
contour.
\begin{figure}
    \centering
\includegraphics[width=1\linewidth]{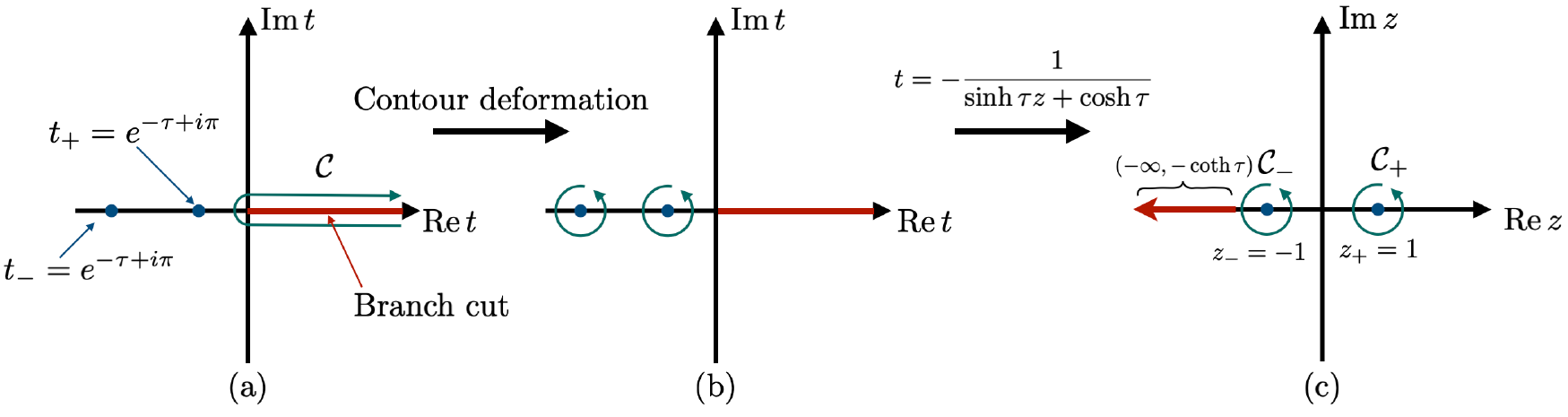}
    \caption{Contour deformation from the $t$-integral to the $z$-integral in equation \eqref{eq:t_integrand_to_z_integrand}.
(a) The integral over $t\in\mathbb{R}_+$, defined as the discontinuity across the branch cut of $t^{2\Delta-1}$, can be deformed to contours (b) around the two poles
$t_+ = e^{-\tau+i\pi}$ and $t_- = e^{\tau-i\pi}$.
Under the map $t=-(\sinh\tau\,z+\cosh\tau)^{-1}$, these poles are sent to the fixed points
$z=\pm1$ of $g_\tau(z)= (\cosh\tau\,z+\sinh\tau)/(\sinh\tau\,z+\cosh\tau)$, leading to (c)the fixed-point contour representation of $G^\pm(\tau)$.
}
    \label{fig:placeholder}
\end{figure}
There is a small contour-theoretic point.  Since $t^{2\Delta-1}$ is
multi-valued, the integral over $t\in\mathbb R_+$ is a boundary value
defined with a branch cut along the positive real $t$-axis.  Equivalently,
one may represent it by the contour that runs just above and below this
cut and takes the corresponding discontinuity.  The map
\eqref{eq:t_z_change} does not send the positive real $t$-axis itself
to the full real line.  For example, for $\tau>0$, it maps
$t\in\mathbb R_+$ to the interval $z\in(-\infty,-\coth\tau)$.

The fixed point contour is obtained by an equivalent deformation.  The
cut contour in the $t$-plane may be deformed to contours enclosing the
two poles of the denominator.  For $\tau>0$, these poles are
\begin{equation}
  t_+ = -e^{-\tau}+i\epsilon=e^{-\tau+i\pi},
  \qquad
  t_- = -e^\tau-i\epsilon=e^{\tau-i\pi}.
\label{eq:t_poles_BD}
\end{equation}
Under the change of variables \eqref{eq:t_z_change} these poles are
mapped to the two fixed points of $g_\tau$:
\begin{equation}
  t_+ \ \to\  z_+=1,
  \qquad
  t_-\  \to\  z_-=-1.
\end{equation}
Equivalently, we have $
  t_+^2
  =
  e^{-2\tau+2\pi i}
  =
  g'_\tau(1)$ and $
  t_-^2
  =
  e^{2\tau-2\pi i}
  =
  g'_\tau(-1)$.
It follows that the positive-time boundary value of the $t$-integral
can be written as a fixed point contour integral
\begin{equation}
  G^+(\tau)
  =
  \int_{{\cal C}_+}
  \frac{dz}{2\pi i}\,
  \frac{g_\tau'(z)^\Delta}
       {z-g_\tau(z)-i\epsilon},
\label{eq:Gplus_from_t_integral}
\end{equation}
where ${\cal C}_+$ denotes the image, in the $z$-plane, of the
deformed contour that encloses the two $t$-plane poles.  For the
opposite time ordering, one takes the complex conjugate boundary value,
which gives
\begin{equation}
  G^-(\tau)
  =
  \int_{{\cal C}_-}
  \frac{dz}{2\pi i}\,
  \frac{g_\tau'(z)^\Delta}
       {z-g_\tau(z)+i\epsilon}.
\label{eq:Gminus_from_t_integral}
\end{equation}

\medskip
\section{More details on the HKLL representation}\label{app:HKLL}
In this appendix we spell out how the representation
(\ref{eq:HKLLbilocal}) is related to the standard two-mode
near-boundary expansion of a massive scalar in dS$_3$
\cite{Xiao:2014uea,Doi:2024nty}.  We define the bilocal 
operator 
\bea
  \mathbb O_{\Delta}(z,\bar z)
  \equiv
  {\cal O}^L_\Delta(\bar z)\,
  {\cal O}^R_\Delta(z),
  \qquad
  (h,\bar h)=(\Delta,\Delta).
\eea
which via the holographic dictionary is placed on the future boundary $\mathscr{I}^+$.
The HKLL-like representation used in the main text \footnote{Here we omit $c_{\Delta}$ in the definition of $K_{\Delta}$ for simplicity.}(\ref{eq:HKLLbilocal}) is
\begin{eqnarray}
  & & \ \ \mathbb O_\Delta(X)
  =
  \int_{\mathscr I^+} d^2z\,
  K_{1-\Delta}(X;z,\bar z)\,
  \mathbb O_{\Delta}(z,\bar z),
\label{eq:HKLL_all_boundary_app}\\[2mm]
 & &   K_{\alpha}(X;z,\bar z)
  =
 e^{-2\pi i\alpha }\left(
  \frac{\eta}
       {\eta^2-(\bar z-\bar x)(z-x)+i\epsilon}
  \right)^{2\alpha}.\label{eq:HKLL_all_boundary_app2}
\end{eqnarray}
In the flat coordinate centered at
$X=(\eta,x,\bar x)$, the future boundary decomposes as
\begin{equation}
  \mathscr I^+
  =
  {\cal D}_+(X)\cup{\cal D}_+(X_A),
\end{equation}
with $
  {\cal D}_+(X):\ |z-x|<\eta$ and ${\cal D}_+(X_A):\ |z-x|>\eta$.
Accordingly,
\bea\label{eq:splitrep}
  & &\hspace{-5mm} \mathbb O_\Delta(X)
  =
  \mathbb O_{\rm in}(X)
  +
  \mathbb O_{\rm out}(X_A),\\[3.5mm]
  \mathbb O_{\rm in}(X) \is 
  \int_{|z-x|<\eta} d^2z\,
  K_{1-\Delta}(X;z,\bar z)\,
  \mathbb O_\Delta(z,\bar z),
\label{eq:Iin_def_app}
\\[1mm]
  \mathbb O_{\rm out}(X_A)
\is
  \int_{|z-x|>\eta} d^2z\,
  K_{1-\Delta}(X;z,\bar z)\,
  \mathbb O_\Delta(z,\bar z),
\label{eq:Iout_def_app}
\eea
Here the branch cut of the kernel $K_{1-\Delta}$ (\ref{eq:HKLL_all_boundary_app2}) sits along $\eta^2-(\bar z-\bar x)(z-x)>0$, i.e., the region ${\cal D}_+(X)$ where the bulk point $X$ is time-like separated from the boundary point $(0, z,\bar{z})$. For the region ${\cal D}_+(X_A)$ space-like separated from the bulk point $X$, we can safely just take $\epsilon\rightarrow 0$ in the expression (\ref{eq:HKLL_all_boundary_app2}) as we do not have the branch cut.
Therefore, $\mathbb O_{\rm in}(X)$ and $\mathbb O_{\rm out}(X_A)$ are
the two lightcone-supported HKLL pieces which isolate, respectively,
one of the two near-boundary modes: $ \mathbb O_\Delta(z,\bar z)$ of the dS$_3$ scalar field associated
with the bulk point $X$ and its antipodal point $X_A$. After acting on the vacuum, this representation corresponds to the bulk local state in dS$_3$ in \cite{Doi:2024nty}.\footnote{In
\cite{Doi:2024nty}, the Euclidean/Bunch--Davies bulk local state is obtained
only after specifying the de Sitter conjugation and the relative phase
between the $\Delta$ and $1-\Delta$ components.  Thus the
identification is not an equality of the raw split expression (\ref{eq:splitrep}) with the bulk local state in
\cite{Doi:2024nty} before fixing this physical pairing convention.  In our
convention the Bunch--Davies branch is selected by the Lorentzian
$i\epsilon$ prescription of the kernels, whereas in
\cite{Doi:2024nty} it is selected by the de Sitter conjugation together
with the CPT-invariant linear combination of the two modes.}

On the other hand, the standard formula of the HKLL representation in dS$_3$, the bulk operator $\mathbb O^{\rm HKLL}_\Delta(X)$ is constructed from the two near-boundary modes via \cite{Xiao:2014uea}: 
\begin{align}
  \mathbb O^{\rm HKLL}_\Delta(X)
  =
  \int_{\strut|z-x|<\eta}\!\!\!\!\!\!\!\! d^2z\,
  K_{1-\Delta}(X;z,\bar z)\,
  \mathbb O_\Delta(z,\bar z)
  -
  \int_{\strut |z-x|<\eta}\!\!\!\!\!\!\!\!\! d^2z\,
  K_{\Delta}(X;z,\bar z)\,
  \widetilde{\mathbb O}_{1-\Delta}(z,\bar z).
\label{eq:HKLL_two_inner_disks_app}
\end{align}
The non-trivial point we make here is that the outer contribution $\mathbb O_{\rm out}(X_A)$ may be rewritten as
an integral over the inner disk, at the price of replacing the boundary
operator by its two-dimensional shadow. 
To demonstrate this, we need the following kernel identity (see also (\ref{eq:I-K-id}) for the corresponding Euclidean arguments):
\begin{align}
\int_{\strut |z'-x|<\eta}\!\!\!\!\!\!\! d^2z'\,
K_{\Delta}(X;z',\bar z')\,
\frac{1}{|z-z'|^{4(1-\Delta)}}
\ =\
&\frac{\pi}{1-2\Delta}e^{-2\pi i\Delta} \Theta(|z-x|-\eta)\,
K_{1-\Delta}(X;z,\bar z).
\label{eq:shadow_kernel_identity_app}
\end{align}
Let us give the proof of
\eqref{eq:shadow_kernel_identity_app}.  Consider
\begin{equation}
  I(z)
  =
  \int_{|z'-x|<\eta} d^2z'\,
  \left(
  \frac{\eta}{\eta^2-|z'-x|^2+i\epsilon}
  \right)^\alpha
  \frac{1}{|z-z'|^{2\beta}} ,
\end{equation}
where $\alpha=2\Delta, \beta=2(1-\Delta)$ with $\alpha+\beta=2$. We assume $\Delta\neq 1$.
By translation, rotation and scaling, set
$z'=x+\eta w,
  z=x+\eta R$ with $R=|z-x|/\eta$.
Then
\begin{equation}
  I(z)
  =
  \eta^{-\beta}J(R) \quad {\rm with}
\quad 
  J(R)
  =
  \int_{|w|<1} d^2w\,
  (1-|w|^2+i\epsilon)^{-\alpha}
  |R-w|^{-2\beta}.
\end{equation}
For $R>1$, use the disk automorphism $w=\frac{\xi+q}{1+q\xi}$ with $q=\frac1R$.
Then
\begin{align}
  1-|w|^2=
  \frac{(1-q^2)(1-|\xi|^2)}
       {|1+q\xi|^2},\quad
  d^2w
  =
  \frac{(1-q^2)^2}
       {|1+q\xi|^4}\,d^2\xi,
\quad 
  |R-w|
  =
  R\,\frac{1-q^2}{|1+q\xi|}.
\end{align}
Using $\alpha+\beta=2$, all factors of $|1+q\xi|$ cancel and one obtains
\begin{align}
  J(R)
  &=
  \frac{R^{-2\beta}}{
  (1-q^2)^{\beta}}
  \int_{|\xi|<1} d^2\xi\,
  (1-|\xi|^2+i\epsilon)^{-\alpha} \ = \ 
  \frac{\pi}{1-\alpha}\,
  (R^2-1)^{-\beta}.
\end{align}
For $0<R<1$, instead use
$w=\frac{\xi+R}{1+R\xi}$.
The same cancellation gives
\begin{align}
  J(R)
  &=
  (1-R^2)^{-\beta}
  \int_{|\xi|<1} d^2\xi\,
  (1-|\xi|^2+i\epsilon)^{-\alpha}
  |\xi|^{-2\beta} \ = \ 
  \pi \frac{
  B(1-\beta,1-\alpha)}{(1-R^2)^{\beta}} = 0.
\end{align}
Since 
$B(1-\beta,1-\alpha) = 0$ for $\alpha+\beta=2$ by analytic continuation.  Thus
\begin{equation}
  J(R)
  =
  \frac{\pi}{1-\alpha}
  \Theta(R-1)
  (R^2-1)^{-\beta}.
\end{equation}
Restoring $\eta$ gives
\begin{equation}
  I(z)
  =
  \frac{\pi}{1-2\Delta}
  \Theta(|z-x|-\eta)
  \left(
  \frac{\eta}{|z-x|^2-\eta^2}
  \right)^{2(1-\Delta)} .
\end{equation}
This precisely
reproduces the kernel $K_{1-\Delta}(X;z,\bar z)$ on the outer region, where the bulk point $X=(\eta, x,\bar{x})$ is spacelike separated from $(0,z,\bar{z})$ on $\mathscr I^+$,
which proves \eqref{eq:shadow_kernel_identity_app}\footnote{The disk integral itself gives the real exterior kernel.  The Lorentzian
\(i\epsilon\) prescription enters only when this result is compared with
the chosen boundary value of the bulk-to-boundary kernel.  With our
convention
$K_\alpha(X;z,\bar z)
  =
  \left(
  \frac{\eta}{\eta^2-|z-x|^2-i\epsilon}
  \right)^{2\alpha}$,
one has on the exterior region
$K_{1-\Delta}(X;z,\bar z)
  =
  e^{-2\pi i\Delta}
  \left(
  \frac{\eta}{|z-x|^2-\eta^2}
  \right)^{2(1-\Delta)} .
$}.

We now define the normalized shadow operator
\begin{equation}
  \widetilde{\mathbb O}_{1-\Delta}(z',\bar z')
  \equiv
  \mathcal{S}_2\mathbb O_\Delta(z',\bar z')
  =
  \frac{2\Delta-1}{\pi}e^{2\pi i\Delta}
  \int_{\mathscr I^+} d^2z\,
  \frac{\mathbb O_\Delta(z,\bar z)}
       {|z-z'|^{4(1-\Delta)}} .
\label{eq:shadow_operator_app}
\end{equation}
Using \eqref{eq:shadow_kernel_identity_app}, the outer HKLL piece becomes
\begin{align}
  \mathbb O_{\rm out}(X_A)
  &=
  \int_{\mathscr I^+} d^2z\,
  \Theta(|z-x|-\eta)
  K_{1-\Delta}(X;z,\bar z)
  \mathbb O_\Delta(z,\bar z)
\nonumber\\
  &=-
  \int_{|z'-x|<\eta} d^2z'\,
  K_{\Delta}(X;z',\bar z')\,
  \widetilde{\mathbb O}_{1-\Delta}(z',\bar z') .
\label{eq:Iout_shadow_app}
\end{align}
Hence the full representation can be written as the sum of two inner-disk as in equation \eqref{eq:HKLL_two_inner_disks_app}.
This is the desired equivalence between the integral expression (\ref{eq:HKLLbilocal}) of the bulk operator over all of $\mathscr{I}^\pm$ and
the HKLL representation involving both the operator and its shadow.
\section{Coarse-grained DSSYK two-point function}
\label{app:dssyk}
In this Appendix we present a few more details of the coarse-grained two-point correlator of the coupled DSSYK models at finite temperature with arbitrary coarse-grain parameter $a\in[1,2]$. This requires taking the exact formulas of double-scaled SYK and analyze the semi-classical limit with given $a$.  The energy $E$ of DSSYK is parameterized by an angle $\theta$
\be
E(\theta)= \frac{2\cos\theta}{\sqrt{\lambda (1-q)}},\quad 
\frac{\dd E}{\dd \theta} =- \frac{2\sin\theta}{\sqrt{\lambda(1-q)}},\quad \theta\in[0,\pi], \qquad q=e^{-\lambda}\in[0,1)
\ee

We wish to evaluate the coarse-grained two point function of two physical operators
\be\label{eq:ds-correlator}
\bra E_{0}|\mathbb{O}_\Delta(t)\mathbb{O}_\Delta(0)|E_{0}\ket_{a}=\int\dd E\thinspace e^{i(E_0-E)t}\thinspace\rho(E)^{a}|\bra E|\Oc_{\Delta_{L}}|E_{0}\ket|^{2}\thinspace|\bra E|\Oc_{\Del_{R}}|E_{0}\ket|^{2},
\ee
The energy density of DSSYK and  matter two-point function involved in \eqref{eq:ds-correlator} are given by
\bea \label{eq:rho-def}
\rho(E)\dd E=\mu(\theta)\dd\theta=\frac{e^{S_0}}{\Gamma_{q}(\pm 2i s_\theta)}\, \dd\theta, 
\qquad \quad s_\theta \equiv \frac{\theta}{\lambda}\\[2mm]
|\bra E|\Oc_{\Delta_{L/R}}|E_{0}\ket|^{2}\thinspace=\mathcal{N}_{\Del_{L/R}}\frac{\Gamma_{q}\bigl(\Del_{L/R}\pm is_0\pm i s_\theta\bigr)}{\Gamma_{q}(2\Del_{L/R})}.
\eea
where $\Gamma_q(x)$ is the q-deformed Gamma function. We have (for small $\lambda$)
\be
\rho(E)^{a}\dd E=\mu(\theta)^{a}\left(\frac{\dd\theta}{\dd E}\right)^{a-1}\dd\theta=\frac{e^{a S_0}}{\Gamma_{q}(\pm 2is_\theta)^{a}}\left(\frac{\lambda}{2\sin\theta}\right)^{a-1}\dd\theta 
\ee
We are interested in the leading semi-classical limit $\lambda\to 0$ of the physical two-point function
\be
\begin{aligned}
& \left\langle E_0\right| \mathbb{O}_\Delta(t) \mathbb{O}_\Delta(0)\left|E_0\right\rangle_{a}  \; =\;  \int_0^\pi \mathrm{d} \theta \,  e^{i\left(E_0-E(\theta)\right) t} \widetilde{G}_{\Delta,a}(\theta_0,\theta) \\[2mm]
\widetilde{G}_{\Delta,a}(\theta_0,\theta) &  =\    e^{a S_0} \mathcal{N}_{\Delta_L} \mathcal{N}_{\Delta_R}\left(\frac{\lambda}{2 \sin \theta}\right)^{a-1}
\frac{\Gamma_q\bigl(\Delta_L \pm i s_0 \pm s_\theta\bigr) \Gamma_q\bigl(\Delta_R \pm is_0 \pm i s_\theta \bigr)}{{\Gamma_q\left(2 \Delta_L\right) \Gamma_q\left(2 \Delta_R\right)} \Gamma_q\left( \pm 2i s_\theta\right)^a}.
\end{aligned}
\ee
Introduce $s_\theta =s_0 + \alpha$ where $s_0 = \theta_0/\lambda$ corresponds to the energy eigenstate $E_0$ which lies in the spectrum, meaning $0<\theta_0<\pi$,   the integral becomes 
\be \label{eq:ds-correlator2}
\begin{aligned}
 \int_{-\frac{\theta_0} \lambda}^{\frac{\pi-\theta_0} \lambda} & \!\!\! \dd\alpha \; e^{i\left(2 \sin \th_0 \right) t \alpha }\left(\frac{\lambda}{2 \sin \left(\theta_0+\lambda \alpha\right)}\right)^{a-1} \\
& \quad \times \frac{\Gamma_q\bigl(\Delta_L \pm (2is_0 +i \alpha)\bigr) \Gamma_q\bigl(\Delta_R \pm(2is_0+i \alpha)\bigl)\Gamma_q\left(\Delta_L \pm i \alpha\right) \Gamma_q\left(\Delta_R \pm i \alpha\right)}{\Gamma_q(2 \Delta_L) \Gamma_q\left(2 \Delta_R\right) \Gamma_q\left( \pm 2i(s_0+ i \alpha)\right)^a} .
\end{aligned}
\ee
The ratio of $\Gamma_q$ functions simplifies dramatically in the $\lambda \to 0$ limit:
\bea
\frac{\Gamma_{q}(\Delta\pm 2is_\theta)}{\Gamma_{q}(\pm 2 is_\theta)}\,\simeq\, \left(\frac{2\sin\theta}{\lambda}\right)^{2\Delta}
,\qquad\ \frac{\Gamma_{q}(\pm 2i(s_{{\nspc}_\theta}+\alpha))}{\Gamma_{q}(\pm 2i s_{{\nspc}_\theta})}\,\simeq\, e^{-(\pi-2\theta)\alpha}
\eea
Plugging the above into \eqref{eq:ds-correlator2}, and keep the leading order in $\lambda$, we find the integral becomes\footnote{We used the fact that $\Gamma_q(x)/\Gamma(x)=1+\mathcal{O}(\lambda)$, for any $x\in\mathcal{O}(\lambda^0)$. This fails if $x\in\mathcal{O}(\lambda^{-1})$}: 
\be
\begin{aligned}
\int_{-\infty}^{\infty}\!\!  \mathrm{d} \alpha\, e^{2(a-1)\left(\pi-2 \theta_0\right) \alpha+i\left(2 \sin \theta_0\right) \alpha t} \frac{\Gamma\left(\Delta_L \pm i \alpha\right) \Gamma\left(\Delta_R \pm i \alpha\right)}{\Gamma\left(2 \Delta_L\right) \Gamma\left(2 \Delta_R\right)}.
\end{aligned}
\ee
Evaluation of this integral leads to the result in Table \ref{tab:results} after introducing the auxiliary parameter $v$ as well as inverse temperature $\beta$ via $\theta_{0}=\frac{\pi}{2}(1+v),$ and $\beta={\pi v}/{\cos\frac{\pi v}{2}}.$


\printbibliography

\end{document}